\documentclass[journal]{IEEEtran}
\usepackage{setspace}

\usepackage{epsfig}
\usepackage{amsmath}
\usepackage{amsfonts}
\usepackage{color}

\begin{document}
\title{{Joint Relay Selection and Power Allocation in Large-Scale MIMO Systems with Untrusted Relays and Passive Eavesdroppers}\thanks{This work was supported by the INSF and contract number 94023761.
A. Kuhestani and  A. Mohammadi are with Department of Electrical Engineering, Amirkabir University of Technology, Tehran, Iran (emails: a.kuhestani@aut.ac.ir; abm125@aut.ac.ir).}\thanks{
M. Mohammadi is with the Faculty of Engineering, Shahrekord University, Shahrekord, Iran (e-mail: m.a.mohammadi@eng.sku.ac.ir).}}
\author{Ali Kuhestani, \IEEEmembership{Student Member,{\hspace{.7mm}}IEEE}, Abbas Mohammadi $Senior~Member,~IEEE$,\\ and Mohammadali Mohammadi, $Member,~IEEE$}
\markboth{}{Ali Abbas}
\maketitle

\markboth{}{Ali Abbas}

\begin{abstract}

In this paper, a joint relay selection and power allocation (JRP) scheme is proposed to enhance the physical layer security  of a cooperative network, where a multiple antennas source communicates with a single-antenna destination in  presence of untrusted relays and passive eavesdroppers (Eves).  The objective is to protect the  data confidentially while concurrently  relying on the untrusted relays as potential Eves to improve both the security and reliability of the network. To realize this objective, we consider cooperative jamming performed by the destination while JRP scheme is implemented. With the aim of maximizing the instantaneous secrecy rate, we derive a new closed-form solution for the optimal power allocation and propose a simple relay selection criterion under two scenarios of {\it non-colluding} Eves (NCE) and {\it colluding} Eves
(CE). For the proposed scheme, a new closed-form expression is derived for the ergodic secrecy rate (ESR)  and the secrecy
outage probability as security metrics, and a new closed-form expression is presented for the average symbol error rate (SER) as a reliability measure over Rayleigh fading channels. We further explicitly characterize the high signal-to-noise ratio slope and power offset of the ESR to highlight the impacts of system parameters on the ESR. In addition, we examine the diversity order of the proposed scheme to reveal the achievable secrecy performance advantage. Finally, the secrecy and reliability diversity-multiplexing tradeoff of the optimized network are provided.  Numerical results highlight that the ESR performance of the proposed JRP scheme for NCE and CE cases is increased with respect to the number of untrustworthy relays.
\end{abstract}


\section{Introduction}
\IEEEPARstart{L}{arge}-scale multiple-input multiple-output (MIMO) system  as a promising solution of the fifth-generation (5G) wireless
communication networks provides significant performance gains in terms of energy saving and spectral efficiency {\cite {rusek}}, {\cite {larsson}}. This new technology deploys simple coherent processing methods, e.g., maximum ratio transmission (MRT) across arrays of hundreds of or even more antennas at base station (BS) and supports tens of or more mobile users (MUs) {\cite {rusek}}--\cite{rusek2}. {An  attractive feature of large-scale MIMO systems is that they offer a significant security improvement compared to a conventional MIMO systems, 
as	with large-scale multiple antennas (LSMA) at the BS, a  narrow directional beam can be radiated toward the desired terminal. Accordingly, the received signal power
at the desired terminal is several orders of magnitude higher than that at any non-coherent passive eavesdropper (Eve) {\cite {rusek2}}\footnote{{It is worth noting that in contrast to passive eavesdropping, an active Eve may cause pilot contamination at the BS and pose more serious security problem to the network {\cite{Ng}}, \cite{IOT2}.}}.} However, the security benefits of LSMA systems are severely hampered in cooperative networks, where the intermediate nodes may be potential Eves {\cite {he1}}.

Owing to the broadcast nature of wireless communication, the information transmission between legitimate users can be simply captured by Eves. Accordingly, physical layer security (PLS) as a promising approach to enhance the confidentiality of wireless communications has attracted a lot of interest {\cite {mukherjee}}, \cite{review1}. Physical layer secure transmission is provisioned by intelligently exploiting the time varying properties of fading channels, instead of relying on conventional cryptographic techniques {\cite {mukherjee}}. Among the proposed PLS solutions, cooperative relaying, cooperative jamming (CJ) and a mixed of these two techniques have recently attracted a great deal of interest {\cite{srt1}}--{\cite{kimHeo}}. 
Cooperative relaying can enhance the PLS through implementing distributed beamforming or  opportunistic relaying (OR) \cite{srt1}, \cite{HMWang}. In contrast to distributed beamforming that suffers from high complexity due to the need of large overhead, OR imposes low system overhead by choosing only one relay which offers the best secrecy performance for the network {\cite{HMWang}}. {As a complementary approach, CJ schemes can also be implemented by the network's nodes to transmit jamming signals towards the Eves in order to degrade the received signal-to-noise ratio (SNR) at the Eves {\cite{HMwang2015}}--{\cite {kimHeo}}.
This goal can be achieved through applying the following three methods: 1) Source-based jamming {\cite{HMwang2015}}, {\cite{phil_jam}}  in which the source transmits a mixed signal carrying the information signal and
the jamming signal,  2) Friendly jammer-based jamming  {\cite{phil_jam}}--{\cite{Debbah}} in which a relay \cite{phil_jam}--{\cite{wang_hanzo}} or an external friendly jammer (FJ)
 {\cite{Debbah}} contributes to provide confidential communication, and 3) Destination-based CJ (DBCJ) {\cite{Petrop}}--{\cite{KCTeh}}, in which the destination itself contributes to degrade the received signal at an external Eve  {\cite{Petrop}}--\cite{kimHeo} or at a helper intermediate node who may act as an Eve \cite {he1}. Among the presented methods, the DBCJ policy can
be implemented simply compared to the first and second methods, where the destination performs self-interference
cancellation with regard to its prior information of jamming
signal. Contrary to DBCJ, for the first and second methods, the destination must be perfectly aware of the pre-defined jamming signal, while it is unknown for the Eves. This {\it a priori} known jamming signal, generated by using some pseudo-random codes or some cryptographic signals, is hard to implement and transfer to the destination confidentially. Transferring this known jamming signal from the source to destination imposes more challenges to the network, when an untrusted relay collaborates for  data transmission.}

From a perspective of security, a trusted relay can friendly assist to protect the confidential
message from being eavesdropped by illegitimate nodes, while an {\it untrusted relay} may intentionally overhear the information signal when relaying. In some communication networks, an untrusted relay may collaborate to provide a reliable communication {\cite {he1}}. This scenario occurs in large-scale wireless systems such as heterogeneous networks, device-to-device (D2D) communications and Internet-of-things (IoT) applications \cite{IOT2}, 
where confidential messages are often retransmitted by multiple intermediate nodes. It is therefore necessary to answer this question that whether exploiting the untrusted relay is still beneficial compared with direct transmission (DT) and if so, what the appropriate relaying strategy should be.

To achieve a positive secrecy rate in untrusted relaying networks, the DBCJ policy was first introduced in {\cite {he1}}. Then several recent works  have investigated the performance of the DBCJ policy in presence of a single {\cite {LWang}}, \cite{Kuh-OPA} or multiple {\cite{sun}}--{\cite{KCTeh}} untrusted relays. Specifically, the secrecy performance of DBCJ with optimal power allocation (OPA) in presence of an untrusted amplify-and-forward (AF) relay  investigated in {\cite{LWang}} and then comprehensively studied in \cite{Kuh-OPA}. Taking into account non-colluding untrusted relaying, the authors in {\cite {sun}}--{\cite{sec-aware}} derived lower bound  expressions for the ergodic  secrecy rate (ESR) performance in the absence {\cite{sun}}, \cite{cioffi} or presence \cite{sec-aware} of source jamming. The researchers in {\cite {sun}}--{\cite{sec-aware}} indicate that increasing the number of untrustworthy relays degrades the ESR in contrast to the case of trustworthy relays. To be specific, the authors in {\cite{cioffi}} found that the diversity order of OR is restricted to unity independent of the number of untrusted relays. All the aforementioned works {\cite{sun}}--\cite{KCTeh} investigated the cooperative untrusted relaying networks in absence of passive Eves and without considering achievable secrecy rate optimization. 

{While the recent literature {\cite{LWang}} presented a solid work
	for OPA in presence of a single untrusted relay, the
	impact of OPA on a more realistic cooperative network with multiple untrusted relays and passive Eves
	has not been studied yet.
	In this paper, motivated by the recent literature on LSMA-based relaying systems {\cite{rusek}}, {\cite{larsson}}, {\cite{IOT2}}, \cite{LWang}, \cite{Chen2}, \cite{Relay}, we present a comprehensive research to improve the PLS of cooperative networks in presence of untrusted relays and passive Eves\footnote{Henceforth, we interchangeably use the terms `passive Eve' and `Eve'.}. Along this line, we consider a new cooperative network consisting of a  large-scale MIMO source, a destination, multiple non-colluding untrusted AF relays {\cite{sun}}--{\cite{KCTeh}}, and multiple non-colluding Eves (NCE) {\cite{HMWang}}, \cite{jefAnd} who hide their existence in the network. In contrast to vast studies on the secrecy performance
	of untrusted relaying networks {\cite {LWang}}--{\cite{KCTeh}}, we take into account joint relay selection and power allocation (JRP) in our network by considering the information leakage in the second phase of transmission. We further investigate a worst-case scenario, where passive Eves may collaborate together to maximize the total received SNR. Toward this end, we consider the scenario that the maximal ratio combining (MRC) is performed over all the received signals from the source and the selected relay 
	to perform more harmful attacks. This event is referred to as colluding Eves (CE) \cite{jefAnd}, \cite{phil_jam_MRC}}. As a benchmark, we next study the conventional DT dispensing with the relays to compare with the proposed JRP transmission policy. In contrast to {\cite{LWang2}}, in which DT policy with transmit antenna selection is analyzed for MIMO wiretap channels, we utilize the simple MRT beamformer {\cite{larsson}} in our proposed DT policy to maximize the received SNR at the intended receiver. Furthermore, different from {\cite{LWang2}} that considered one multiple antennas Eve, we consider two cases of NCE and CE. Therefore, our system model and the related analysis are completely different compared with {\cite{LWang2}}.

 The main contributions of the paper are summarized as follows:
\begin{enumerate}
\item We develop a JRP scheme to maximize the instantaneous secrecy rate of the network for both NCE and CE cases. Our findings highlight that the large-scale MIMO
approach as a powerful mathematical tool offers a new simple relay selection criterion. The proposed criterion only requires the channel state information (CSI) of the relays-destination links. 
\item Based on the proposed JRP scheme,  new closed-form expressions are derived for the probability of positive secrecy rate and the ESR of NCE and CE cases over Rayleigh fading channels. Furthermore, new compact expressions are derived for the asymptotic ESR.  Our asymptotic results highlight that the probability of positive secrecy transmission tends to one for the optimized JRP scheme as the number of relays $K$ grows. Moreover, we find that the ESR performance improves as $K$ increases. We further
 characterize  the high SNR slope and power offset of the ESR to highlight the impact of system parameters on the ESR performance. 
\item  For the proposed JRP scheme,  new closed-form expressions are derived for the  secrecy outage probability (SOP) of both NCE and CE cases over Rayleigh fading channels. In order to shed insights into the system performance, new simple expressions are derived for the SOP in the high SNR regime.  Our asymptotic results highlight that the proposed JRP scheme enjoys the diversity order of $K$. Next, the secrecy diversity-multiplexing trade-off (DMT) is examined to express the trade-off between the error probability and the data rate of the proposed JRP scheme.
\item For  DT policy, a new tight lower
bound is derived for the ESR of NCE and CE cases. We also derive a new closed-form
expression for the SOP of those cases. Our results show that the  high SNR slope and the secrecy diversity order of this transmission policy are zero. 
  \item To illustrate the reliability of the network, we calculate the average symbol error rate (SER) as the performance measure. New exact and asymptotic closed-form expressions are derived for the SER performance of the proposed JRP scheme over Rayleigh fading channels. Our results highlight that the proposed scheme offers the diversity order of $K$. We further determine the reliability DMT of the proposed scheme.
\end{enumerate}

The remainder of this paper is organized as follows. In Section II we present the system model and preliminaries.
Next, in Section III, the relay selection criterion is introduced and  the related OPA is evaluated. Performance metrics are evaluated in Section IV. Numerical results are presented in Section V,
followed by conclusions in Section VI.

{\it Notation}: We use bold lower case letters to denote vectors. ${\bf I}_N$ and ${\bf 0}_{N\times 1}$ denote the Identity matrix and the zeros matrix, respectively. $\|.\|$ and $(.)^H$ denote the Euclidean norm and conjugate transpose operator, respectively;
$\mathbb{E}_x\{\cdot\}$ stands for the expectation over the random variable x; $\Pr(\cdot)$ denotes the probability; $f_X(\cdot)$ and $F_X(\cdot)$ denote the probability density function (pdf) and cumulative distribution function (cdf) of the random variable (r.v.) $X$, respectively;  $\mathcal{CN} (\mu, \sigma^2)$ denotes a circularly symmetric complex Gaussian
r.v. with mean $\mu$ and variance $\sigma^2$; $\mathrm{Ei}(x)$, $Q(x)$ and $\Psi(x)$ are  the exponential integral {\cite[Section (8.21)]{table}, the Q-function {\cite[Section (8.25)]{table} and the psi function {\cite[Section (8.25)]{table}, respectively. $[\cdot]^+=\max\{0,x\}$ and $\max$ stands for the maximum value. $\lceil x \rceil$ is the smallest integer that is larger than or equal to $x$.

\begin{figure}[t] 
  \begin{center}{\hspace{-10mm}}
    \includegraphics[width=3.6in]{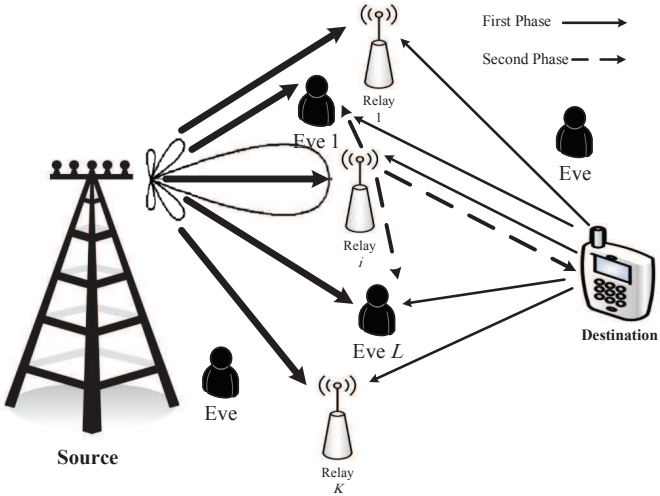} 
    \caption{Secure transmission in a cooperative network in  presence of multiple untrusted relays and multiple passive Eves. Under DBCJ policy, relay $i$ is selected to amplify the source's signal. The solid and dashed lines denote the first and second phases of transmission, respectively.}
  \label{fig_system}\end{center}
\end{figure}
\section{System Model}

We consider a cooperative LSMA-based network consisting of one multiple antennas  source equipped with an array of $N_\mathrm{s}$ antennas, one single-antenna destination, $K$ single-antenna untrusted AF relays and $L$ single-antenna passive Eves\footnote{The presence of multiple Eves in a wireless network may be a realistic scenario in which the malicious nodes act as harmful attackers to the authorized terminals {\cite{IOT2}}, \cite{phil_jam_MRC}.} as depicted in Fig. \ref{fig_system}. {The untrusted
relays in our network are so-called semi-trusted, i.e., they are trusted at the service level
while they are untrusted at the data level\footnote{{The service level trust
	conveys this concept that the accurate CSI of the communication links can be
	given to the source through relay's cooperation, and the relay retransmits the received information signal
	toward the destination. However, this collaboration
	is untrustworthy at the data level, i.e., the relay may decipher the confidential information from their received signal.}} {\cite{he1}}.} All the passive Eves are randomly located around the source, the relay nodes and the destination, and they hide their existence in the network. We mention that, in this paper, the term ``malicious node'' includes both the untrusted relays and passive Eves. We further assume that each node operates in a half-duplex mode. Based on time division duplexing (TDD) operation, the source obtains downlink CSI through uplink training. Then
the acquired CSI is used to generate the low implementation complexity MRT precoding matrix \cite{Chen2}.

We assume that the untrusted relays extract the information signal solely based on its own observation and they adopt selection combining (SC) {\cite{sec-aware}}, while relying on the Eves' behavior, the following two eavesdropping scenarios are considered in our work:
\begin{enumerate}
	\item {\it Non-colluding eavesdroppers case:} In this case, each passive Eve individually overhears the information signal without any collaboration with other Eves. For this case,  we assume SC is applied at the Eves. As will be validated via numerical examples, the presented analysis can still be utilized for the case of MRC at the malicious nodes.
	\item {\it Colluding eavesdroppers case:} In this case, all the passive Eves can connect to a  data center to share their information, leading to extract more information. For this case,  we assume MRC technique is applied at the Eves to enhance the intercept probability. This case can be considered as a worst-case scenario from the  security viewpoint \cite{jefAnd}.  
\end{enumerate}

 Some additional assumptions and definitions are as follows:
\begin{itemize}
  \item Complex Gaussian channel vector from the source to the malicious node $l$: ${\bf h}_{{sl}}\sim \mathcal{CN}({\bf 0}_{{N_\mathrm{s}}\times 1},\mu_{{sl}}{\bf {I}}_{N_\mathrm{s}})$.
  \item Complex Gaussian channel vector from the source to the destination: ${\bf h}_{{sd}}\sim \mathcal{CN}({\bf 0}_{{N_\mathrm{s}}\times 1},\mu_{{sd}}{\bf {I}}_{N_\mathrm{s}})$.
  \item Complex Gaussian channel from the selected relay $i$ to the destination: $ {h}_{{id}}\sim \mathcal{CN}( 0,\mu_{{id}})$.
  \item Complex Gaussian channel from the selected relay $i$ to the malicious node $l$: $ {h}_{{il}}\sim \mathcal{CN}( 0,\mu_{{il}})$.
  \item All the channels satisfy the reciprocity theorem {\cite {sun}}.
  \item The additive white Gaussian noise (AWGN) at each receiver $n_m$ ,$m\in\{{i,l,d}\}$, is a zero-mean complex Gaussian r.v. with variance $N_0$.
  \item The total transmit power of each phase is limited by $P$. One practical reason is the impact of co-channel interference between adjacent networks which should be taken into account in network design.
  \item $\rho=\frac{P}{N_0}$ is the transmit SNR of the system.
  \item $\gamma_{{si}}=\rho\|{\bf h}_{{si}}\|^2$, $\gamma_{{id}}=\rho|{ h}_{id}|^2$, $\gamma_{{il}}=\rho|{ h}_{il}|^2$, $\gamma_{ld}=\rho |h_{ld}|^2$, $\gamma_{{sd}}=\rho\|{\bf h}_{{sd}}\|^2$ and  $\gamma_{null,l}^m=\rho |\frac{{\bf h}_{sm}^H}{|{\bf h}_{{sm}}|}{\bf h}_{sl}|^2$, $m \in \{i,d\}$.
  \item $\bar \gamma_{si}=\rho \mu_{si}$,  $\bar \gamma_{{id}}=\rho \mu_{{id}}$, $\bar \gamma_{sl}=\rho \mu_{sl}$, $\bar \gamma_{ld}=\rho \mu_{ld}$ and $\bar \gamma_{{sd}}=\rho \mu_{{sd}}$.
\end{itemize}
In the following, we describe two transmission policies adopted in this paper to establish the PLS in our considered network; namely DBCJ policy and DT policy. 

\subsection{DBCJ Policy}
{In some communication networks,
	the direct channel gain between the source and destination may be so weak when the source and destination are located far apart or within heavily shadowed areas. In this condition, cooperative relaying offers a promising solution. This scenario has been extensively implemented in most previous works {\cite{HMWang}}, \cite{phil_jam}--{\cite{wang_hanzo}}, \cite{kimHeo}, where the source and relays pertain to a group, while the destination
	and passive Eves are placed in another group. Some networks including this scenario are the mobile ad-hoc
	networks (MANETs) \cite{srt1},  the LTE
	cellular systems  \cite{srt1}, {\cite{wang_hanzo}} and the IoT network \cite{IOT2}, \cite{IOT1}.}
The aim of DBCJ policy is to deteriorate the received signal at the malicious nodes and to not allow them to decipher the information signal. 

{Now, we proceed to highlight the necessity of adopting jamming signal in our relay-aided network. Toward this end and for the first step, we investigate the conventional AF relaying scheme in which CJ is not utilized for data transmission.
Since the selected relay $i$ itself is a curious node, the instantaneous secrecy rate is given by {\cite{sun}}
	\begin{align}
	R_s^{\mathrm{conv}}=\Big[\frac{1}{2}\log_2\Big(1+\frac{\gamma_{si}\gamma_{id}}{1+\gamma_{si}+\gamma_{id}}\Big)-\frac{1}{2}\log_2(1+\gamma_{si})\Big]^+.
	\end{align} 
	According to the fact that  $\frac{xy}{1+x+y}<\min(x,y)\leq x$ {\cite{sun}}, one can easily conclude that  $R_s^{\mathrm{conv}}=0$. This result indicates that the achievable secrecy rate of the conventional untrusted relaying scheme without employing CJ is zero. Therefore, a jamming signal must be
	sent to degrade the received signal at the untrusted relay. 
	 In this paper, due to the simplicity of DBCJ policy compared with source-based jamming and FJ-based jamming, we adopt the DBCJ policy to provide perfect secure transmission. It is worth noting that the analysis in this paper can be simply extended to the case of utilizing an external FJ {\cite{Debbah}}.}}

 Under DBCJ policy, since the nodes operate
 in a half-duplex mode, the direct link between the source and destination is missed. The whole transmission is performed based on a time division
 multiple-access (TDMA) protocol including the broadcast phase (first phase) and the relaying phase (second phase). During the first phase, while the source transmits the information signal with power $\lambda P$, the destination concurrently radiates the artificial noise with power $(1-\lambda)P$, where $\lambda \in (0,1)$ is the power allocation factor. It is worth noting that this power allocation scheme
provides insights for the power allocation of the source and destination. This power allocation approach has been exploited in several works for both performance analysis
and network optimization design  {\cite{HMWang}}, {\cite{LWang}}--{\cite{sun}}, \cite{sec-aware}. During the second phase, the selected relay normalizes its received signal and forwards it with power $P$. Finally, the destination decodes the source information by subtracting the self-interference signal. Note that due to the  broadcast nature of wireless communication, the signal transmitted in both the phases could be eavesdropped. {In this paper, contrary to most recent literature in multiple untrusted relays {\cite{sun}}--{\cite{KCTeh}} that ignored the information leakage in the second phase of transmission, we consider a more realistic scenario. In our considered scenario, the malicious nodes intercept the transmissions of both the source and the selected relay and then try to capture the information.}

Denoting ${x}_\mathrm{s}$ and ${x_\mathrm{d}}$ as the information signal and the jamming signal, respectively, the received signal at the malicious node $l$ ($l\in \{1,...,K+L\}$) in first phase can be expressed as
\begin{eqnarray}\label{yy_r}
y_{l}^{(1)}=\sqrt{\lambda P} {\bf w}^H{\bf{h}}_{sl}{x}_{s}+\sqrt{(1-\lambda)P}  h_{ld} { x}_{d}+n_{l},
\end{eqnarray}
where ${\bf w}=\frac{{\bf h}_{{si}}}{{\Arrowvert {\bf h}_{{{si}}}\Arrowvert}}$ represents the MRT beamformer at the source and $i$ ($i \in \{1,...,K\}$) indicates the index of the selected relay $\mathrm{R}_i$. In the second phase, the relay amplifies its received signal by an amplification factor of
\begin{eqnarray}
G=\sqrt{\frac{P}{\lambda P {{\Arrowvert {\bf h}_{{{si}}}\Arrowvert}}^2+(1-\lambda)P|h_{{id}}|^2+N_0}},
\end{eqnarray}
 and broadcasts the message $x_{{i}}=Gy_{{i}}$. During the second phase, all the malicious nodes hear the signal diffused by $\mathrm{R}_i$. Thus, the received signal at the malicious node $l$ ($l \neq i$) can be expressed as
\begin{eqnarray}\label{yy_r2}
y_{{l}}^{(2)}=Gh_{il}y_i^{(1)}+n_{{l}}.
\end{eqnarray}
Furthermore, at the destination, after performing self-interference cancellation, the corresponding signal is given by
\begin{eqnarray}\label{yy_d}
{y}_{D}{\hspace{-3mm}}&=&{\hspace{-3mm}}\sqrt{\lambda P}G {{\Arrowvert {\bf h}_{{{si}}}\Arrowvert}}h_{{id}}{x}_{s}+Gh_{{id}}n_{{j}} + n_{d}.
\end{eqnarray}
Based on (\ref {yy_r}), the received signal-to-interference-and-noise-ratio (SINR) at the selected relay $i$ and the SINR at the malicious node $l$ during the first phase of transmission are respectively, given by
\begin{eqnarray}\label{SINR_r}
\gamma_{{i}}=\frac{\lambda \gamma_{{si}}}{(1-\lambda)\gamma_{{id}}+1}~~~\mathrm{and}~~~
\gamma_{{l}}^{(1)}=\frac{\lambda \gamma_{null,l}^i }{(1-\lambda)\gamma_{ld}+1}.
\end{eqnarray}
By substituting (\ref {yy_r}) into (\ref {yy_r2}), the received SINR at the malicious node $l$ during the second phase can be expressed as 
\begin{eqnarray}\label{SINR_r2}
\gamma_l^{(2)}=\frac{\lambda \gamma_{si}\gamma_{il}}{\lambda \gamma_{si}+(1+(1-\lambda)\gamma_{id})(1+\gamma_{il})}.
\end{eqnarray}
Moreover, by invoking (\ref {yy_d}), the SINR at the destination can be obtained as
\begin{eqnarray}\label{SINR_d}
\gamma_{{D}_i}=\frac{\lambda \gamma_{{si}} \gamma_{{id}}}{\lambda \gamma_{{si}}+(2-\lambda){\gamma_{{id}}}+1}.
\end{eqnarray}

The instantaneous secrecy rate corresponding to the selected relay $i$ can be expressed as {\cite {he1}}
\begin{eqnarray}\label{rate_exact}
R_s^{(i)}\hspace{-3mm}&=&\hspace{-3mm}\frac{1}{2}\Big[\log_2(1+\gamma_{{D}_i})-\log_2(1+\gamma_E)\Big]^+,
\end{eqnarray}
where $\gamma_E$ is the amount of information leaked to the malicious nodes. In case of NCE, the information leakage is given by
\begin{eqnarray}\label{leakage_NCE}
\gamma_{E_i}^{NCE}=\max_{1\leq l\leq K+L, l\neq i}\Big\{\gamma_{{i}},\gamma_{{l}}^{(1)}, \gamma_{{l}}^{(2)}\Big\}.
\end{eqnarray}
Moreover, the information leakage for CE case is given by \cite{jefAnd}, {\cite{phil_jam_MRC}}
\begin{eqnarray}\label{leakage_CE}
\gamma_{E_i}^{CE}=\max\Big\{\hspace{-1mm}\underbrace{\max_{1 \leq l \leq K, l\neq i }\{\gamma_i,\gamma_l^{(1)},\gamma_l^{(2)}\}}_{\mathrm{Information~leakage~ 1}},\hspace{-2mm}\underbrace{\sum_{l=1}^{L}(\gamma_{{l}}^{(1)}+ \gamma_{{l}}^{(2)})}_{\mathrm{Information~ leakage~2}}\hspace{-2mm}\Big\},
\end{eqnarray}
where information leakages 1 and 2 describe the information leakage for untrusted relays and passive Eves, respectively.

{ {\it Remark 1: }In practical cooperative networks, e.g., an ultra-dense heterogeneous wireless network, both trusted and untrusted relays exist. In such a network, to combat the
	potential security attacks from untrusted relays, reliable relay authentication techniques are implemented, i.e.,
the network needs to discriminate between the trusted and untrusted relays and then adopts the suitable secure transmission policy. The  
	authentication can be accomplished by traditional key-based cryptographic methods or new physical layer
	authentication (PLA) methods \cite{Authen}, {\cite{hanzoA}}. A novel technique to implement the PLA is the multi-attribute multi-observation (MAMO) technique {\cite{hanzoA}} in which the joint verification of the received signal strength indicator (RSSI) and the hardware imperfections of the transceivers are exploited to authenticate the relay nodes. PLA can also be implemented based on this feature that the device-dependent
	hardware impairment in-phase/quadrature (I/Q) imbalance related to the reception
	and transmission of the relays is unique \cite{Authen}.}

\subsection{DT Policy}In DT policy, when the source transmits the information signal to the destination, all the relays and the Eves listen. Accordingly, the received SNR at the malicious node $l$ and the SNR at the destination are respectively, represented by
 \begin{eqnarray}\label{g_DT}
\gamma_{l}{\hspace{-.5mm}}={\hspace{-.52mm}}\gamma_{null,l}^d\stackrel {(a)}{\approx}\frac{\rho \hat{h}_{dl}}{N_\mathrm{s} \mu_{sd}}~~~\mathrm{and}~~~
\gamma_{D}{\hspace{-.52mm}}={\hspace{-.52mm}}\rho \|{\bf h}_{{sd}}\|^2\stackrel {(a)}{\approx} N_\mathrm{s} \overline{\gamma}_{sd},
\end{eqnarray}
where $\hat{h}_{dl}\stackrel{\Delta}{=}|{\bf h}^H_{{sd}}{\bf h}_{{sl}}|^2$. In (\ref{g_DT}), $(a)$ follows from the law of large numbers {\cite {papul}} due to large $N_s$. Applying the Lindeberg-Levy central limit theorem {\cite {papul}}, the r.v. $\hat{h}_{dl}$ can be approximated as $\mathcal{CN}(0,N_{\mathrm{s}}\mu_{{sd}}\mu_{{sl}})$. It is worth pointing out that $\hat{h}_{dl}$ is very well approximated as a Gaussian r.v. even for small $N_{\mathrm{s}}$ {\cite {guass1}}. Therefore, $\gamma_{null,l}^d$ can be considered as an exponential r.v. with mean $\overline{\gamma}_{null,l}^d=\rho \mu_{{sl}}$. 

For DT policy, the information leakage for NCE and CE cases can be respectively, expressed as \cite{jefAnd},  {\cite{phil_jam_MRC}}
\begin{align}\label{leakage_NCE_DT}
\gamma_{E}^{NCE}&=\max_{1 \leq l \leq K+L}\{\gamma_l\}~~~\mathrm{and}~~~\gamma_{E}^{CE}=\max_{1 \leq l \leq K}\Big\{\gamma_l,\sum_{j=1}^{L}{\gamma_j}\Big\}.
\end{align}

\section{Joint Relay Selection and Power Allocation}
In this section, we first introduce the optimal relay selection scheme. Then motivated by the LSMA {\cite{rusek}}, {\cite{larsson}}, our proposed JRP scheme is presented for both NCE and CE cases.

Let $i^{\star}$ denote the index of the best selected relay. A relay that maximizes the instantaneous secrecy rate is selected as
\begin{eqnarray}\label{selection_ex}
{i^{\star}}=\mathrm{arg} \max_{1\leq i\leq K}R_s^{(i)}.
\end{eqnarray}
The optimal relay selection criterion (\ref {selection_ex}) requires the CSI of both hops and all inter-relay channels. Therefore, it is rather complicated to be implemented in practice, especially when the number of source antennas and the number of relays are large. To alleviate this issue, motivated by LSMA at the source, we will propose a simple relay selection criterion for both NCE and CE cases.

{\subsection{Non-colluding Eavesdroppers}}
For NCE case, based on (\ref{leakage_NCE}), we need to find the maximum received SINR at the malicious nodes. Due to deploying an LSMA at the source and by leveraging the Cauchy-Schwarz inequality, $\gamma_l^{(1)}$ in (\ref {SINR_r}) can be upper bounded by
\begin{eqnarray}\label{SNR_r_1}
\gamma_l^{(1)}=\frac{\lambda_{NCE} \gamma_{null,l}^i}{(1-\lambda_{NCE})\gamma_{ld}+1}<\gamma_i,
\end{eqnarray}
where $\lambda_{NCE}$ represents the power allocation factor related to NCE case. Furthermore, $\gamma_l^{(2)}$ in (\ref{SINR_r2}) can be upper bounded by
\begin{eqnarray}\label{SNR_r_2}
\gamma_l^{(2)}{\hspace{-3mm}}&<&{\hspace{-3mm}}\frac{\lambda_{NCE} \gamma_{si}\gamma_{il}}{(1+(1-\lambda_{NCE})\gamma_{id})(1+\gamma_{il})}\nonumber\\
{\hspace{-3mm}}&<&{\hspace{-3mm}}\frac{\lambda_{NCE} \gamma_{si}}{1+(1-\lambda_{NCE})\gamma_{id}+1}= \gamma_i.
\end{eqnarray}
According to (\ref {SINR_r}), (\ref {SNR_r_1}) and (\ref {SNR_r_2}), the information leakage in (\ref {leakage_NCE}) is simplified as $\gamma_E^{NCE}={\gamma}_{{i}}$. As such, the instantaneous secrecy rate in  (\ref {rate_exact}) is simplified as
\begin{eqnarray}\label{rate_exact_@}
R_s^{(i)}=\frac{1}{2}\Big[\log_2\Big(\frac{1+\gamma_{{D}_i}}{1+{\gamma}_{{i}}}\Big)\Big]^+.
\end{eqnarray}
Let $\phi(\lambda_{NCE})\stackrel {\triangle}{=}\frac{1+\gamma_{{D}_i}}{1+{{\gamma}}_{{i}}}$. Notably $\phi(0) \geq 1$. Therefore, for OPA $\phi(\lambda_{NCE}^{\star})\geq 1$, where $\lambda_{NCE}^\star$ is the OPA factor corresponding to  NCE case. Therefore, the operator $[\cdot]^+$ in (\ref {rate_exact_@}) can be dropped. For NCE case, we have the following key
result.\\

{\it Proposition 1}: For a large number of antennas at the source, the function $\phi(\lambda_{NCE})=\frac{1+\gamma_{{D}_i}}{1+{{\gamma}}_{{i}}}$ is a quasiconcave function of ${\lambda_{NCE}}$ in the feasible set and the optimal solution is given by
\begin{eqnarray}\label{solution111}
\lambda_{NCE}^{\star}=\frac{\sqrt {2} \gamma_{id}}{\gamma_{si}}.
\end{eqnarray}

{\it Proof}: By substituting the expressions (\ref {SINR_r}) and (\ref {SINR_d}) into (\ref {rate_exact_@}), forming the function ${\phi (\lambda_{NCE})}$, and then taking the first derivative of it with respect to $\lambda_{NCE}$, we obtain
\begin{align}\label{derivative}
&\phi'(\lambda_{NCE})=\nonumber\\
&-{\frac {\gamma_{si} \left(  C_1 {\lambda_{NCE}}^{2}+ C_2 \lambda_{NCE}+C_3 \right) }{ \gamma_{id}\left( \lambda_{NCE}\,\frac{\gamma_{si}}{\gamma_{id}}-
\lambda_{NCE}+1 \right) ^{2} \left( \lambda_{NCE}\,\frac{\gamma_{si}}{\gamma_{id}}-\lambda_{NCE}+2 \right) ^{2}}},
\end{align}
where $C_1=(\frac{\gamma_{si}}{\gamma_{id}}-1) \Big( ( {\it {\gamma_{{id}}}}+1
 ) \frac{\gamma_{si}}{\gamma_{id}}+2\,{\it {\gamma_{{id}}}}-1 \Big)$, $C_2=\left( 4\,{\it {\gamma_{{id}}}}+4
\,\frac{\gamma_{si}}{\gamma_{id}}-4 \right)$ and
$C_3=-2\,{\it {\gamma_{{id}}}}+4$. For a large number of antennas at the source, $\gamma_{si} \gg \gamma_{id}$, the coefficients are simplified as $C_1=(\gamma_{{id}}+1)\frac{\gamma_{si}^2}{\gamma_{id}^2}$, $C_2=4\frac{\gamma_{si}}{\gamma_{id}}$ and $C_3=-2\gamma_{{id}}+4$. By solving ${\phi'(\lambda_{NCE})}=0$, the single feasible solution is obtained as (\ref {solution111}). Since ${\phi'(1)}=-\frac{\gamma_{si}^3(1+\gamma_{{id}})}{4\gamma_{id}^3}<0$, we conclude that  ${\phi (\lambda_{NCE})}$ is a quasiconcave function in the feasible set.$~~\square$

 Thanks to the LSMA at the source, by applying the law of large numbers {\cite {papul}}, we have $\gamma_{{si}}\approx N_\mathrm{s} \overline{\gamma}_{{si}}$. Therefore, the optimal solution (\ref {solution111}) that requires the CSI of both hops can be further simplified as
\begin{eqnarray}\label{solution222}
{\lambda}_{NCE}^{\star}\approx \frac{\sqrt 2 |{h_{{id}}}|^2}{N_{\mathrm{s}}~\mu_{{si}}}.
\end{eqnarray}
The proposed OPA factor in (\ref {solution222}) requires only the CSI of the selected relay-destination link and the statistical mean of the source-selected relay link. 

 By substituting (\ref {solution222}) into (\ref {SINR_r}) and (\ref {SINR_d}), we obtain
\begin{eqnarray}\label{gamma_E_D_NCE}
\gamma_{E_i}^{NCE}={{\gamma}_{{i}}}\approx \sqrt{2}~~~~~~~~\mathrm{and}~~~~~~{\gamma_{{D}_i}^{NCE}}\approx \frac{{\gamma_{{id}}}}{1+{\sqrt 2}}.
\end{eqnarray}
Therefore, by using (\ref{gamma_E_D_NCE}), the instantaneous secrecy rate in (\ref {rate_exact}) can be rewritten as
\begin{eqnarray}\label{cap2}
R_{s}^{(i)}=\frac{1}{2}\log_2\Big(\frac{1+\gamma_{{D}_{i}}^{CE}}{1+\sqrt{2}}\Big).
\end{eqnarray}
The interesting result in (\ref {cap2}) indicates that the instantaneous secrecy rate only depends on the transmit SNR and the relay-destination link. As a consequence, the high complexity relay criterion (\ref {selection_ex}) can be approximated as
\begin{eqnarray}\label{selection}
i^{\star}&=&\mathrm{arg} \max_{1\leq i \leq K}\gamma_{{D}_i}^{CE}\nonumber\\
&=&\mathrm{arg} \max_{1\leq i \leq K} |{h}_{{id}}|^2.
\end{eqnarray}
The proposed relay selection criterion in (\ref {selection}) requires only the relays-destination channel gains and hence, enjoys from low complexity and energy consumption. Furthermore, the relay selection scheme in (\ref {selection}) can be easily implemented using the distributed timer technique in {\cite {bletsas}}.

In practice, the proposed JRP scheme can be implemented as follows: Before data transmission, the relays are scheduled to transmit pilot symbols {\cite{Chen2}}. Using the pilots, the source and the destination can estimate their channels. Then according to (\ref {selection}), the destination computes the strongest link between itself and the relays and hence, the relay index $i^{\star}$ is selected. Afterward, the destination broadcasts $i^{\star}$ and pilot symbols to estimate the destination-relay $i^{\star}$ link. Then the relay $i^{\star}$ forwards a quantized version of the estimated destination-relay $i^{\star}$ to the source ($l$ bits). Finally, both the source and  destination tune their optimal transmit power to start communication. Accordingly, the overall overhead required for broadcasting is equal to $l+\lceil \log_2K \rceil$ bits.\\

{\subsection{Colluding Eavesdroppers}}
 For CE case, the Eves jointly try to decode the information signal based on MRC processing and hence, the amount of overheard information increases. To tackle this problem, more power should be dedicated to the destination to confuse the Eves compared to $\lambda_{NCE}^{\star}$ in (\ref {solution111}) and less power to the source to transmit the information signal. Based on this, $\lambda_{CE}^{\star}<\lambda_{NCE}^{\star}\ll 1$, where $\lambda_{CE}^{\star}$ denotes the OPA factor for CE case. Therefore, $\gamma_l^{(2)}$ in (\ref{SINR_r2}) can be approximated as
\begin{align}\label{gamma_2_CE}
\gamma_l^{(2)}\hspace{-1mm}\approx\hspace{-1mm} \frac{\lambda_{CE} \gamma_{si} \gamma_{il}}{(1+(1-\lambda_{CE})\gamma_{id})(1+\gamma_{{il}})}\hspace{-1mm}\hspace{-.5mm}\stackrel{(a)}{\approx}\hspace{-1mm} \frac{\lambda_{CE} \gamma_{si}}{(1-\lambda_{CE})\gamma_{id}+1}\hspace{-.5mm}=\hspace{-.5mm}\gamma_i,
\end{align}
where $(a)$ follows from the high SNR assumption. The interesting result in (\ref{gamma_2_CE}) expresses that the amount of information leakage to the malicious nodes in the second phase of transmission is approximately the same as the amount of information leaked to the selected relay in the first phase. Based on this new result and given this fact that, the received signal by the malicious nodes in the second phase of transmission is a degraded version of the emitted signal by the selected relay, $\gamma_{E_i}^{CE}$ in (\ref{leakage_CE}) is changed to
	\begin{align}\label{gamma_CE_total}
	\gamma_{E_i}^{CE}=\gamma_i+\sum_{l=1}^L \gamma_l^{(1)}.
	\end{align}{For the sake of tractability, according to the fact that $\lambda_{CE}^* \ll 1$, $\gamma_i$ and $\gamma_l^{(1)}$ in (\ref{SINR_r}) can be rewritten as
\begin{align}\label{gamma_CE_new}
\gamma_i\hspace{-1mm} \approx\hspace{-1mm} \frac{\lambda_{CE} \gamma_{si}}{(1-\lambda_{CE})(\gamma_{id}+1)}, ~~ \gamma_l^{(1)}\hspace{-1mm}\approx\hspace{-1mm} \frac{\lambda_{CE} \gamma_{null,l}^i}{(1-\lambda_{CE}) (\gamma_{ld}+1)}.
\end{align}
By substituting (\ref{gamma_CE_new}) into (\ref{gamma_CE_total}), we obtain	
\begin{align}\label{gamma_E_CE}
\gamma_{E_i}^{CE}=\frac{\lambda_{CE} \Delta}{1-\lambda_{CE}},
\end{align}
where $\Delta=\frac{\gamma_{si}}{\gamma_{id}+1}+\sum_{l=1}^L\frac{\gamma_{null,l}^i}{\gamma_{ld}+1}$. 
Following the same steps as in Proposition 1, at the high SNR regime, the OPA factor corresponding to CE case can be obtained as
\begin{align}\label{sol_CE}
\lambda_{CE}^*=\sqrt{\frac{2 \gamma_{id}}{\gamma_{si}\Delta}}.
\end{align}
We can conclude from (\ref{sol_CE}) that by increasing the number of Eves, the most amount of the total power is assigned to the destination to inject jamming signal.}  

{\it Remark 2:} In practice, often may not be feasible to achieve the CSI of passive Eves. Based on this practical issue, we consider a scenario, where only the second order
statistics related to the Eves' are available which is a common assumption in the literature {\cite{HMWang}}, {\cite{krikid}}, \cite{wang_hanzo}, \cite{kimHeo}. Furthermore, for mathematical simplicity, we assume that the relaying and eavesdropping channels are independent
and identically distributed, i.e., for relaying channels with $1 \leq 1\leq K$, we consider $\mu_{sl}=\mu_{sr}$ and  $\mu_{ld}=\mu_{rd}$, and for eavesdropping channels with $1\leq l\leq L$, we consider $\mu_{sl}=\mu_{se}$ and $\mu_{ld}=\mu_{ed}$ \cite{srt1}, {\cite{krikid}}, {\cite{wang_hanzo}}.

Here, we consider the case with large number of passive Eves. As will be observed in numerical examples, the analysis are valid even for moderate number of Eves.  Based on the law of large numbers and since the first and the second hops are independent, we have
\begin{align}\label{asym_eve}
\sum_{l=1}^L\frac{\gamma_{null,l}^i}{\gamma_{ld}+1}&\approx L~\mathbb{E} \Big\{\gamma_{null,l}^i\Big\} \mathbb{E} \Big\{\frac{1}{\gamma_{ld}+1}\Big\}\nonumber\\
&=-\frac{L \mu_{se}~e^{\frac{1}{\rho \mu_{ed}}}}{\mu_{ed}} \mathrm{Ei}\Big(-\frac{1}{\rho \mu_{ed}}\Big)\stackrel{\Delta}{=}\hat{\theta},
\end{align}
where we used \cite[Eq. (3.352.4)]{table}. By substituting (\ref{asym_eve}) into $\lambda_{CE}^{\star}$ in (\ref{sol_CE}) and then substituting  into (\ref{gamma_E_CE}) and (\ref{SINR_d}), we obtain
\begin{align}\label{gamma_CE}
\gamma_{E_{i}}^{CE}\approx \sqrt{2+2\frac{\gamma_{id}}{\gamma_{si}}\hat{\theta}}~~~\mathrm{and} ~~~\gamma_{D_i}^{CE}=\frac{\gamma_{id}}{1+\sqrt{2+2\frac{\gamma_{id}}{\gamma_{si}}\hat{\theta}}}.
\end{align}
By substituting $\gamma_{E_i}^{CE}$ and $\gamma_{D_i}^{CE}$ in (\ref{gamma_CE}) into (\ref{rate_exact}),  the instantaneous secrecy rate is obtained. Accordingly, we find that the optimal relay selection scheme corresponding to the maximum achievable secrecy rate needs the excessive implementational overhead. The optimal relay selection based on both hops is out of the scope of this paper. As such, we adopt
the suboptimal simple relay selection scheme proposed for NCE case in (\ref{selection}). 
We
mention that since the second hop has a dominant impact on quantifying the received SNR at the destination, the relay selection criterion in (\ref{selection}) can be considered as a suboptimal criterion to enhance the secrecy rate. Moreover, in case of very LSMA at the source, the suboptimal relay selection criterion in (\ref{selection}) will be the near-optimal one. 

For the selected relay $i^*$ with the relay selection criterion in (\ref{selection}) and with very LSMA at the source, we can use the well-known approximation of minimum mean square error (MMSE) estimator as follows \cite{rusek}, \cite{Kay} \begin{align}\label{app_devide}
\frac{\gamma_{i^*d}}{\gamma_{si^*}}&\approx \mathbb{E}\Big\{\frac{\gamma_{i^*d}}{\gamma_{si^*}}\Big\}\stackrel{(a)}{=}\mathbb{E}\Big\{\gamma_{i^*d}\Big\}\mathbb{E}\Big\{\frac{1}{\gamma_{si^*}}\Big\}\nonumber\\
&\stackrel{(b)}{=}\frac{ \mu_{rd}(\Psi(K+1)+\varrho)}{(N_s-1)\mu_{sr}}\stackrel{\triangle}{=}\eta,
\end{align}
where $\mathcal{\varrho}=0.577$ {\cite {table}} is Euler's constant. In (\ref{app_devide}), $(a)$ follows from the fact that the two-hops are independent and $(b)$ follows from evaluating $\mathbb{E}\{\gamma_{i^*d}\}$ and $\mathbb{E}\{\frac{1}{\gamma_{si^*}}\}$. For the first case, $\mathbb{E}\{\gamma_{i^*d}\}$, we have
\begin{align}
\mathbb{E}\Big\{\gamma_{i^*d}\Big\}&=\int_0^{\infty}\Big(1-F_{\gamma_{i^*d}}(x)\Big) dx\nonumber\\
&={\rho \mu_{rd}} \Big(\Psi(K+1)+\varrho\Big),
\end{align}
where the last equality follows from using the cdf of $\gamma_{i^*d}$ which can be obtained based on order statistics {\cite {papul}} 
\begin{align}\label{theorem2}
F_{\gamma_{{i^{\star} d}}}(\gamma){\hspace{-0mm}}&={\hspace{-0mm}}\prod_{i=1}^K \Big[1-\exp{(-\frac{\gamma}{\overline {\gamma}_{{id}}})}\Big].
\\\label{CDF}
&=1+\sum_{n=1}^{2^K-1}(-1)^{|\mathfrak{u}_n|}\exp{\Big(-\sum_{i \in \mathfrak{u}_n}\frac{\gamma}{\overline {\gamma}_{{id}}}\Big)},
\end{align}
where $|\mathfrak{u}_n|$ denotes the cardinality of the $n$-th non-empty subcollection of the $K$ relays, and the last expression follows from applying the binomial expansion theorem. For the second case $\mathbb{E}\{\frac{1}{\gamma_{si^*}}\}$, we used {\it Lemma 2.9} in \cite{randomMat}
\begin{align}
\mathbb{E}\Big\{\frac{1}{\gamma_{si^*}}\Big\}=\frac{1}{(N_s-1)\rho \mu_{sr}}.
\end{align}
Therefore, by substituting (\ref{app_devide}) into (\ref{gamma_CE}), we arrive at
\begin{align}\label{gamma_CE_final}
\gamma_{E_{i^*}}^{CE}\approx \sqrt{2+2\eta\hat{\theta}}~~~\mathrm{and} ~~~\gamma_{D_{i^*}}^{CE}=\frac{\gamma_{i^*d}}{1+\sqrt{2+2\eta\hat{\theta}}}.
\end{align}
We conclude from (\ref{gamma_CE_final}) that the constant value $\gamma_{E_{i^*}}^{CE}$  increases by increasing the number of Eves in the network. By equipping the source with very LSMA $N_s\rightarrow \infty$, this information leakage tends to $\sqrt{2}$ the same as NCE case.

Comparing (\ref{gamma_E_D_NCE}) for NCE and (\ref{gamma_CE_final}) for CE, we can define the new parameter $C$ to integrate the performance analysis for NCE and CE cases. Based on this definition, we have
\begin{align}\label{gamma_combine}
\gamma_{E_i^*}\approx \sqrt{{2(1+C)}}~~~\mathrm{and}~~~\gamma_{D_{i^*}}\approx\frac{\gamma_{{i^*d}}}{1+\sqrt{2(1+C)}},
\end{align}
where $C=0$ for NCE case and $C=\eta \hat{\theta}$ for CE case.\\

{\it Remark 3:} Both the network power optimization  and performance analysis presented in this paper, can be routinely extended to the scenario that the untrusted relays and passive Eves collaboratively decode the information signal based on MRC technique.\\

{\it Remark 4:} According to the results in (\ref{SNR_r_1}) and (\ref{gamma_2_CE}), the received SNR at the malicious nodes in the second phase of transmission is more than that one in the first phase of transmission. Therefore, when SC is adopted at the malicious nodes, and all of them cooperate to decode the information signal, the network power optimization  and performance analysis are similar to the NCE case.  \\

\section{Performance Analysis}
In this section, we derive new closed-form expressions for the probability of positive secrecy rate, the ESR and the SOP as the metrics of security and the SER as the reliability measure. Both the proposed JRP scheme (which is based on DBCJ policy) and the DT policy are studied.

\subsection{Probability of Positive Secrecy Rate}
In this section, we proceed to derive the probability of positive secrecy rate for the proposed DBCJ and DT policies.

{\bf { {DBCJ Policy}}}: By substituting (\ref{gamma_combine}) into (\ref{rate_exact}) and using the proposed relay selection criterion in (\ref {selection}), the probability of positive secrecy rate can be expressed as
\begin{align}\label{psr1}
P_{pos}^{\mathrm{DBCJ}}&=\Pr\Big\{R_s^{(i^{\star})}>0\Big\}\nonumber\\
&=\Pr\Big\{\gamma_{{i^{\star} d}}>{2(1+C)+\sqrt{2(1+C)}}\Big\}\nonumber\\
&=1-F_{\gamma_{{i^{\star} d}}}\Big(2(1+C)+\sqrt{2(1+C)}\Big)\nonumber\\
&\stackrel {(a)}{=}\hspace{-1mm}1{\hspace{-1mm}}-{\hspace{-1mm}}\prod_{i=1}^K \Big[1{\hspace{-1mm}}-{\hspace{-1mm}}\exp{\Big(-\frac{2(1+C)+\sqrt{2(1+C)}}{ \overline {\gamma}_{{id}}}\Big)}\Big],
\end{align}
where $(a)$ follows from substituting (\ref{theorem2}). {It can be concluded from (\ref{psr1}) that, the proposed JRP is not efficient when the average transmit SNR of the second hop ${\overline {\gamma}_{{id}}}$ (which is a function of the transmit SNR $\rho$ and the distance-dependent channel gain  $\mu_{id}$) is low. This observation is not surprising, because when the DBCJ policy is adopted, due to the half-duplex operation of the nodes, the direct link between the source and destination is vanished. As such, the destination only relies on the second hop to receive the information signal form the source. Consequently, the reliability of confidential communication is degraded when the relays are far from the destination or the transmit SNR is low.} In the high SNR regime, (\ref {psr1}) is simplified as
\begin{eqnarray}\label{high_snr_pos}
P_{pos}^\mathrm{DBCJ}=1-\Big(2(1+C)+\sqrt{2(1+C)}\Big)^K \prod_{i=1}^K \frac{1}{\overline {\gamma}_{{id}}}.
\end{eqnarray}
We deduce from (\ref {high_snr_pos}) that the probability of positive secrecy rate approaches one as the transmit SNR or the number of relays increases. The reason is that as $K$ grows, the occurrence probability of a stronger channel between relays and the destination increases and thus, $P_{pos}^\mathrm{DBCJ}$ approaches one.\\

{ \bf {DT Policy}}: For this policy, we will study the NCE case and CE case separately, below.

\subsubsection{{NCE Case}} In this case, by combining (\ref {rate_exact}), (\ref {leakage_NCE}) and (\ref {g_DT}), the probability of positive secrecy rate can be obtained as
\begin{align}\label{P_pos_dt2}
P_{pos}^{\mathrm{DT, NCE}}&=\Pr\Big\{\log_2\Big(1+N_\mathrm{s} \overline{\gamma}_{sd}\Big)
\nonumber\\
&-\log_2\Big(1+\max_{1 \leq l \leq K+L}\gamma_{null,l}^d\Big){\hspace{-1mm}}>{\hspace{-1mm}}0\Big\}\nonumber\\
&\approx \prod_{l=1}^{K+L}\Big(1-e^{-\frac{N_\mathrm{s} \mu_{sd}}{\mu_{{sl}}}}\Big)^l,
\end{align}
where the last expression follows from the fact that $\gamma_{null,l}^d$ can be approximated as an exponential r.v., as mentioned in Section II-B. This result  reveals that $P_{pos}^{\mathrm{DT,NCE}} \rightarrow 0$ as the number of malicious nodes goes to infinity. Furthermore, one can obtain $P_{pos}^{\mathrm{DT,NCE}} \rightarrow 1$ as $N_\mathrm{s} \rightarrow \infty$, which is an intuitive observation, since  by increasing $N_\mathrm{s}$, the source with an LSMA focuses the transmission energy toward the direction of the selected relay. Hence, the strength of the received signals at the malicious nodes will be low enough. Based on this, the malicious nodes fail to extract the information. {Furthermore, in contrast to DBCJ policy that the confidential communication is solely
	dependent on the second hop channel, expression (\ref{P_pos_dt2}) illustrates that  since the probability of positive secrecy rate for DT policy is independent of the transmit SNR, the DT can offer secure transmission even for low transmit SNRs by deploying large number of source antennas $N_s$.}

\subsubsection{{CE Case}} As mentioned in (\ref{leakage_NCE_DT}), in this case   $\gamma_E^{CE}=\max_{1 \leq l \leq K}\{\gamma_l, \sum_{j=1}^{L}{\gamma_j}\}.$ To obtain the probability of positive secrecy rate, we first find the cdf of $\gamma_E^{CE}$. Using order statistics {\cite {papul}}, the cdf of $\gamma_E^R\stackrel{\Delta}{=}\max_{1\leq l\leq K}\{\gamma_l\}$ is given by \cite{srt1}, \cite{papul}
\begin{eqnarray}\label{CDF_R}
F_{\gamma_E^R}(\gamma)=1+\sum_{n=1}^{2^K-1}(-1)^{|\mathfrak{u}_n|}\exp{\Big(-\sum_{l \in \mathfrak{u}_n}\frac{\gamma}{\overline {\gamma}_{{sl}}}\Big)}.
\end{eqnarray}The following lemma from \cite{Golds} helps to obtain the cdf of $\gamma_E^{Eve}\stackrel{\Delta}{=}\sum_{j=1}^{L}\gamma_j$ in (\ref{leakage_NCE_DT}).\\

{\it Lemma 1}: Let $\{X_i\}_{i=1}^n$, $n>1$, be independent exponential r.vs with distinct  averages $\mu_i$. Then the cdf  of their sum $W={X_1+X_2+...+X_n}$ is given by
	\begin{align}
	F_W(x)=\sum_{j=1}^{n}\Gamma({\mu}_j) \Big(1-e^{-\frac{x}{\mu_j}}\Big), ~x>0.
	\end{align}
where $\Gamma({\mu}_j)\stackrel{\Delta}{=}\Big(\prod_{i=1}^n \frac{1}{\mu_i}\Big) \frac{\mu_j}{\prod_{k \neq j , k=1}^n(\frac{1}{\mu_k}-\frac{1}{\mu_j})}$.

Using (\ref{CDF_R}) and leveraging Lemma 1, the cdf of $\gamma_E^{CE}$ in (\ref{leakage_NCE_DT}) can be expressed as
	\begin{align}\label{cdf_gamma_E_CE}
	&F_{\gamma_E^{CE}}(\gamma)=F_{\gamma_E^R}(\gamma) F_{\gamma_E^{Eve}}(\gamma)\nonumber\\
	&={\hspace{-1mm}}\Big(\hspace{-.7mm} 1\hspace{-1mm}+\hspace{-2mm}\sum_{n=1}^{2^K-1}(-1)^{|\mathfrak{u}_n|}\exp{(-\sum_{l \in \mathfrak{u}_n}\frac{\gamma}{\overline {\gamma}_{{sl}}})}\Big)\hspace{-.6mm}\Big(\sum_{j=1}^{L}\Gamma(\bar{\gamma}_{sj}) (1-e^{-\frac{x}{\bar{\gamma}_{sj}}})\Big).	\end{align}
	Hence, the probability of positive secrecy rate for DT policy under the presence of CE can be expressed as
	\begin{align}\label{Ppos_DT_NCE0}
	P_{pos}^{\mathrm{DT,CE}}=F_{\gamma_E^{CE}}(N_\mathrm{s} \overline{\gamma}_{sd}),
	\end{align}
	where the cdf of $\gamma_E^{CE}$ is in (\ref{cdf_gamma_E_CE}). From (\ref{Ppos_DT_NCE0}), we observe that when $N_s \rightarrow \infty$, the secure communication is established, while for large number of malicious nodes, the secure transmission is harmed.

\subsection{Ergodic Secrecy Rate}{\vspace{-.5mm}}
The ESR as a useful secrecy metric characterizes the average of the achievable instantaneous rate difference between the legitimate channel and the wiretap channel.
In the following, we derive  new accurate closed-form expressions for the ESR of both DBCJ based and DT policies.\\

{\bf { {DBCJ Policy}}}:
By substituting (\ref{gamma_combine}) into (\ref{rate_exact}) and using the relay selection criterion in (\ref {selection}), the ESR can be written as
\begin{align}\label{CapT1}
\overline {R_{s}^{\mathrm{DBCJ}}}&=\frac{1}{2\ln 2}\int_0^{\infty}\ln\Big(1+\frac{ \gamma}{1+\sqrt {2(1+C)}}\Big)f_{\gamma_{{{i^{\star}}d}}}(\gamma) {d}\gamma\nonumber\\
&-\frac{1}{2}\log_2\Big(1+\sqrt{{2}(1+C)}\Big)\nonumber\\
&\stackrel {(a)}{=}\frac{1}{2\ln 2}\int_0^{\infty}\frac{\Big[1-F_{\gamma_{{{i^{\star}}d}}}(\gamma)\Big]}{\gamma+1+\sqrt{2(1+C)}} ~\mathrm{d}\gamma\nonumber\\ &-\frac{1}{2}\log_2\Big(1+\sqrt{2(1+C)}\Big),
\end{align}
where $(a)$ follows from using  integration by parts. To obtain a closed-form solution for the ESR, by substituting (\ref {CDF}) into (\ref {CapT1}) and using \cite[Eq. (3.352.4)]{table}, we get
\begin{align}\label{CapT_fin}
&\overline {R_{s}^{\mathrm{DBCJ}}}=\frac{1}{2\ln 2}\sum_{n=1}^{2^K-1}(-1)^{|\mathfrak{u}_n|}{e}^{\sum_{i \in \mathfrak{u}_n}\frac{1+\sqrt{2(1+C)}}{{\overline{\gamma}}_{{id}}}}\nonumber\\
&\times \mathrm{Ei} \Big(-\sum_{i \in \mathfrak{u}_n}\frac{1+\sqrt{2(1+C)}}{{\overline{\gamma}}_{{id}}}\Big)-\frac{1}{2}\log_2(1+\sqrt{2(1+C)}).
\end{align}
We remark that the ESR in (\ref {CapT_fin}) is explicitly characterized by the average channel gains between the relays and destination, and the transmit SNR of the network. As observed, increasing the number of antennas at the source $N_s$ has no impact on the ESR performance. 
For a single-relay network without any passive Eve, the ESR in (\ref {CapT_fin}) is simplified as
\begin{align}\label{CapT000}
\hspace{-1mm}{\overline {R_{s}^{\mathrm{DBCJ}}}}=-\frac{1}{2\ln 2}{e}^{\frac{1+\sqrt{2}}{ \overline {\gamma}_{1d}}}\mathrm{Ei}\Big(-\frac{1+\sqrt{2}}{\overline {\gamma}_{1d}}\Big)-\frac{1}{2}\log_2(1+\sqrt{2}),
\end{align}
which was derived in {\cite{LWang}}, {\cite{Kuh-OPA}}. As such, the works {\cite{LWang}}, {\cite{Kuh-OPA}} can be considered as a special case of our work.

We can further calculate the ESR performance in the high SNR regime when $\rho\rightarrow\infty$ by applying the general asymptotic form given by {\cite {LWang2}}
\begin{eqnarray}\label{asymp}
\overline{R_s^\infty}=S_{\infty}\Big(\log_2 \rho-L_{\infty}\Big),
\end{eqnarray}where $S_{\infty}$ is the high SNR slope in bits/s/Hz/ (3 dB) and $L_{\infty}$ is the high SNR power offset in 3 dB units. These parameters are two key performance factors that
explicitly examine the ESR performance at the high SNR regime which are defined respectively, as \cite{LWang2}
\begin{eqnarray}\label{power_slope}
S_{\infty}=\lim_{\rho \rightarrow \infty}\frac{\overline{R_s^{\infty}}}{\log_2 \rho}~~~\mathrm{and}~~~L_{\infty}=\lim_{\rho \rightarrow \infty}\Big(\log_2 \rho-\frac{\overline{R_s^{\infty}}}{S_{\infty}}\Big).
\end{eqnarray}
We mention that the
high SNR slope is also recognized as the maximum multiplexing gain or the number of degrees of freedom {\cite{LWang2}}. Based on (\ref{gamma_combine}), in the high SNR regime with $\rho \rightarrow \infty$, we
have $\ln(1+\gamma_{D_i})\approx \ln(\gamma_{D_i})$. As
such, the ESR of the proposed JRP scheme can be represented by
\begin{align}\label{rate_asymp}
\overline{R_s^{\infty}}=\frac{1}{2 \ln 2}\mathbb{E}\Big\{\ln (\gamma_{i^{\star}d})\Big\}
-\log_2\Big(1+\sqrt{2(1+C)}\Big).
\end{align}
By taking the derivative of (\ref{CDF}) with respect to $\gamma$ to obtain the pdf of $\gamma_{i^*d}$, and then applying \cite[Eq. (4.331.1)]{table}, the asymptotic ESR can be obtained as
\begin{align}\label{rate-asymp-comp}
\overline{R_s^{\infty}}&=\frac{1}{2 \ln 2}\Big(	\sum_{n=1}^{2^K-1}(-1)^{|\mathfrak{u}_n|}\Big[\ln(\sum_{i \in \mathfrak{u}_n}\frac{1}{\overline {\gamma}_{{id}}})+\mathcal{\varrho}\Big]\Big)\nonumber\\
&-\log_2\Big(1+\sqrt{2(1+C)}\Big).
\end{align}Using \eqref{power_slope}, the high SNR slope is given by
\begin{eqnarray}\label{slope}
S_{\infty}=\frac{1}{2},
\end{eqnarray}
where we used the fact that $\sum_{n=1}^{2^K-1}(-1)^{|\mathfrak{u}_n|}=-1$. Expression (\ref{slope}) highlights that the number of source antennas $N_\mathrm{s}$ and the presence of collaborative eavesdropping have no impact on the ESR slope.

Furthermore, the high SNR power offset is derived as
\begin{align}\label{offset}
L_{\infty}&=-\sum_{n=1}^{2^K-1}(-1)^{|\mathfrak{u}_n|}\log_2(\sum_{i \in \mathfrak{u}_n}\frac{1}{\mu_{{id}}})
+\frac{\mathcal{\varrho}}{\ln 2}\nonumber\\
&+2\log_2\Big(1+\sqrt{2(1+C)}\Big),
\end{align}
where a decrease in the power offset corresponds to an increase in the ESR. Expression \eqref{offset} characterizes the impacts of $\mu_{id}$, the number of relays $K$ and the number of passive Eves $L$ on the ESR and shows that the power offset is independent of $N_\mathrm{s}$. As expected, we find that increasing the number of passive Eves increases the ESR power offset which corresponds to a decrease in the ESR.\\

{\bf { {DT Policy}}}: In the following, we investigate the ESR performance of DT policy for both NCE and CE cases.
\subsubsection{{NCE Case}}
In this case, by substituting $\gamma_E^{NCE}$ in (\ref{leakage_NCE_DT}) and $\gamma_{D}$ in (\ref {g_DT}) into (\ref {rate_exact}), we have
\begin{align}\label{DT_esr_LB}
\overline{R_s^\mathrm{DT,NCE}}{\hspace{-0mm}}&={\hspace{-0mm}}\mathbb{E}\Big\{\Big[\log_2\Big(1+ N_\mathrm{s} \overline{\gamma}_{sd}\Big)\nonumber\\
&-\log_2\Big(1+\max_{1 \leq l \leq K+L}\gamma_{null,l}^d\Big)\Big]^+\Big\}\\
&\geq\Big[\log_2\Big(1+N_\mathrm{s} {\overline{\gamma}}_{sd}\Big)\nonumber\\
&-\mathbb{E}\Big\{\log_2\Big(1+\max_{1 \leq l \leq K+L}\gamma_{null,l}^d\Big)\Big\}\Big]^+{\stackrel{\mathrm{\triangle}}{=}}~\overline{R_{s,LB}^\mathrm{DT,NCE}},\label{DT_esr_ex} \nonumber\\
\end{align}
 where the inequality follows from the fact $\mathbb{E}\Big\{\max\{x,y\}\Big\} \geq \max \Big\{ \mathbb{E}\{x\},\mathbb{E}\{y\}\Big\}$ {\cite {sun}}. Following the same steps presented to derive the ESR of DBCJ policy, (\ref {DT_esr_ex}) is given by
\begin{eqnarray}\label{dt_final}
\overline{R_{s,LB}^\mathrm{DT,NCE}}{\hspace{-3mm}}&=&{\hspace{-3mm}}\Big[\log_2\Big(1+N_\mathrm{s} {\overline{\gamma}}_{sd}\Big)\nonumber\\
&-&{\hspace{-3mm}}
\frac{1}{\ln 2}\hspace{-1mm}\sum_{n=1}^{2^{K+L}-1}(-1)^{|{v}_n|}{e}^{\sum_{i \in {v}_n}\frac{1}{{\overline{\gamma}}_{{sj}}}}\mathrm{Ei} \Big(-\sum_{i \in {v}_n}\frac{1}{{\overline{\gamma}}_{{sj}}}\Big)\Big]^+,\nonumber\\
\end{eqnarray}
where $|v_n|$ denotes the cardinality of the $n$-th non-empty subcollection of the $K+L$ nodes. We remark that at the high SNR regime, the instantaneous secrecy rate (\ref {DT_esr_LB}) is further simplified as
\begin{eqnarray}\label{high_SNR_DT}
R_s^{DT,NCE}{\hspace{-1mm}}={\hspace{-1mm}}\Big[2 \log_2 \Big( N_\mathrm{s} \mu_{sd}\Big){\hspace{-1mm}}-{\hspace{-1mm}}\log_2\Big( \max_{1\leq l \leq K+L} \hat{h}_{ld}\Big)\Big]^+.
\end{eqnarray}
Expression (\ref{high_SNR_DT}) states that the instantaneous secrecy rate of DT is independent of the transmit SNR, i.e., a secrecy rate ceiling appears when the transmit SNR increases. In other words, the high SNR slope of DT policy is zero. We conclude that, unlike the DBCJ policy, the DT cannot achieve high secure transmission rates.  Furthermore, as observed from (\ref {high_SNR_DT}), the ESR decreases by increasing the number of malicious nodes while the ESR increases as $N_\mathrm{s}$ grows. \\

\subsubsection{{CE Case}}  In this case, by using $\gamma_E^{CE}$ in (\ref{leakage_NCE_DT}), a tight lower bound for the ESR of CE case can be obtained as
\begin{align}\label{DT_ESR_CE}
\overline{R_{s,LB}^\mathrm{DT,CE}}&=\Big[\log_2\Big(1+N_\mathrm{s} {\overline{\gamma}}_{sd}\Big)
-\mathbb{E}\Big\{\log_2\Big(1+\gamma_E^{CE}\Big)\Big\}\Big]^+ \nonumber\\
&=\Big[\log_2\Big(1+N_\mathrm{s} {\overline{\gamma}}_{sd}\Big)\hspace{-1.2mm}+\hspace{-1mm}\sum_{j=1}^{L}\Gamma(\bar{\gamma}_{sj})e^{\frac{1}{\bar{\gamma}_{sj}}} \mathrm{Ei} \Big(-\frac{1}{ \bar{\gamma}_{sj}}\Big)\nonumber\\
&-\sum_{n=1}^{2^K-1}\sum_{j=1}^{L}(-1)^{|\mathfrak{u}_n|} \Gamma(\bar{\gamma}_{sj})e^{\sum_{l \in \mathfrak{u}_n}\frac{1}{ \overline {\gamma}_{{sl}}}} \mathrm{Ei}(-\frac{1}{ \bar{\gamma}_{sl}})\nonumber\\
&+\sum_{n=1}^{2^K-1}\hspace{-1mm}\sum_{j=1}^{L}(-1)^{|\mathfrak{u}_n|} \Gamma(\bar{\gamma}_{sj})e^{\frac{1}{\bar{\gamma}_{sj}}+\sum_{l \in \mathfrak{u}_n}\frac{1}{\overline {\gamma}_{{sl}}}}\nonumber\\
&\times\mathrm{Ei}\Big(-\frac{1}{\bar{\gamma}_{sj}}-{\sum_{l \in \mathfrak{u}_n}\frac{1}{\overline {\gamma}_{{sl}}}}\Big)\Big]^+,
\end{align}
where we used the cdf of $\gamma_E^{CE}$ in (\ref{cdf_gamma_E_CE}) and \cite[Eq. (3.352.4)]{table}. Similar to NCE case, a secrecy rate ceiling appears in the high SNR regime which limits the secrecy performance of DT policy.

\subsection{{ {Secrecy Outage Probability}}}
  The overall SOP denoted by $P_{\mathrm{out}}$ is defined as the probability that a system with the instantaneous secrecy rate $R_s$ is unable to support the target transmission rate $R_t$, i.e., $P_{\mathrm{out}}=\Pr\Big\{R_s<R_t\Big\}.$\\

  {\bf DBCJ Policy}: By Substituting (\ref{gamma_combine}) into the SOP definition, we obtain
  \begin{align}\label{SOP}
  P_{\mathrm{out}}^{\mathrm{DBCJ}}&=\Pr\Big\{\gamma_{{{i^{\star}}d}}<\widetilde{R}_t\Big\}\nonumber\\
  &=F_{\gamma_{{i^{\star} d}}}\Big(\widetilde{R}_t\Big)=\prod_{i=1}^K \Big[1-\exp{\Big(-\frac{\widehat{R}_t}{\overline {\gamma}_{{id}}}\Big)}\Big],
  \end{align}
  where $\widetilde{R}_t=\Big(1+\sqrt {2(1+C)}\Big)\Big({2^{2R_t}(1+\sqrt{2(1+C)})-1}\Big).$ To derive (\ref{SOP}), we used the cdf of $\gamma_{i^*d}$ in (\ref{CDF}). Expression (\ref{SOP}) indicates that the SOP approaches zero as the transmit SNR $\rho \rightarrow \infty$. For a single-relay case and without passive Eve, the overall SOP is simplified as
    \begin{eqnarray}\label{SOP1}
  P_{\mathrm{out}}^\mathrm{DBCJ}=1-\exp{\Big(-\frac{(1+\sqrt {2})(2^{2R_t}(1+\sqrt{2})-1)}{\overline {\gamma}_{{1d}}}\Big)},
  \end{eqnarray}
  which was derived in our previous work {\cite{Kuh-OPA}}. Therefore, this new work extends the recent work {\cite{Kuh-OPA}}.
  
    Now, we look into the high
  SNR regime and investigate the diversity order.  
  In the high SNR regime, the closed-form expression in (\ref{SOP}) can be written as
    \begin{align}\label{SOP_ap}
  P_\mathrm{out}^\mathrm{DBCJ}\approx  (\widetilde{R}_t)^K \prod_{i=1}^{K}\frac{1}{\overline {\gamma}_{{id}}}.
  \end{align}
By inspecting (\ref {SOP_ap}), we interestingly find that the diversity order of the  system equals to the number of untrusted relays $K$. {To justify this new result intuitively, we can say that in the considered LSMA-based network, under applying the OPA, the first hop channel becomes deterministic and only the second hop contributes for signal transmission. Increasing the number of untrusted relays, actually increases the probability of having a stronger link in the second hop. Therefore, the secrecy diversity order of the system increases which in turn decreases the secrecy outage probability.}\\

{\bf DT Policy:} To study the SOP of DT policy, we have two cases of NCE and CE.

\subsubsection{{NCE Case}} By substituting $\gamma_E^{NCE}$ in (\ref{leakage_NCE_DT}) and $\gamma_{D}$ in (\ref {g_DT}) into (\ref {rate_exact}), and then formulating the SOP, we obtain
\begin{align}\label {sop_dt}
P_{\mathrm{\mathrm{out}}}^{\mathrm{DT,NCE}}&=\Pr\Big\{\log_2\Big(\frac{1+N_\mathrm{s} \overline{\gamma}_{sd}}{1+ \max_{1\leq l \leq K+L}\gamma_{null,l}^d}\Big)<R_t\Big\}\nonumber\\
\hspace{-0mm}&=1{\hspace{-1mm}}-{\hspace{-1mm}}\Pr\Big\{\max_{1\leq l \leq K+L}\gamma_{null,l}^d <\frac{1+N_\mathrm{s} \overline{\gamma}_{sd}}{2^{R_t}}-1\Big\}\nonumber\\
&=1{\hspace{-1mm}}-{\hspace{-1mm}}\prod_{l=1}^{K+L}\Big[1-\exp\Big(-\frac{1}{\overline{\gamma}_{sl}}(\frac{1+N_\mathrm{s} \overline{\gamma}_{sd}}{2^{R_t}}-1)\Big)\Big].\nonumber\\
\end{align}
The expression (\ref {sop_dt}) indicates that $P_{\mathrm{out}}^{\mathrm{DT,NCE}}\rightarrow 1$ as the number of malicious nodes grows. {One can easily conclude from (\ref {sop_dt}) that an error floor occurs at the high SNR regime. This means that the secrecy diversity order of MRT-based  DT policy is zero. This result is the same as the results in {\cite{LWang2}}, where the authors considered transmit antenna selection at the transmitter 
and receive generalized selection combining at the receiver.} Moreover, we can conclude $P_{\mathrm{out}}^{\mathrm{DT,NCE}}\rightarrow 0$ as $N_\mathrm{s} \rightarrow \infty$.\\

\subsubsection{{CE Case}} In this case, we obtain
\begin{align}\label{sop_dt_CE}
P_{\mathrm{\mathrm{out}}}^{\mathrm{DT,CE}}&=\Pr\Big\{\log_2\Big(\frac{1+N_\mathrm{s} \overline{\gamma}_{sd}}{1+ \gamma_E^{CE}}\Big)<R_t\Big\}\nonumber\\
&=1-F_{\gamma_E^{CE}}\Big(\frac{1+N_\mathrm{s} \overline{\gamma}_{sd}}{2^{R_t}}-1\Big),
\end{align} where $F_{\gamma_E^{CE}}(\gamma)$ is in (\ref{cdf_gamma_E_CE}). Expression (\ref{sop_dt_CE}) states that similar to NCE case, the SOP of CE case tends to a constant in the high SNR regime. Furthermore, we find that  the SOP of CE case approaches zero as $N_\mathrm{s} \rightarrow \infty$.

\subsection{{ {Average Symbol Error Rate}}}

In this subsection, we evaluate the SER of the proposed JRP scheme. The SER of the DT scheme is available in the literature {\cite[Sec. 3]{Golds} and is omitted for brevity.

The instantaneous SER of coherent modulation is in the form of $P_s=\alpha_mQ(\sqrt{\beta_m\gamma_D})$, where $\gamma_D$ represents the received SNR, and $\alpha_m$ and $\beta_m$ depend on the modulation type. Specifically, for rectangular $M$-ary quadrature
amplitude modulation, $\alpha_m=4(1-\frac{1}{\sqrt M})$ and $\beta_m=\frac{3}{M-1}$. Moreover, for $M$-ary phase-shift keying ($M\geq 4$), $\alpha_m=2$ and $\beta_m=2\sin^2\frac{\pi}{M}$ {\cite {Golds}}.\\

Using the average SER of coherent modulation as {\cite {Golds}}
\begin{align}\label{Theorem}
\overline{P_s} = \frac{\alpha_m}{\sqrt{2\pi}} \int_0^\infty{F_{\gamma_D}\Big(\frac{t^2}{\beta_m}\Big)\exp(-\frac{t^2}{2})}{\mathrm{d}}t,
\end{align}
the average SER of the proposed JRP system using (\ref{theorem2}) can be expressed as
\begin{align}\label{SER00}
{\overline{P_s^{\mathrm{DBCJ}}}} &=\frac{\alpha_m}{\sqrt{2\pi}} \int_0^\infty\hspace{-2mm}\exp(-\frac{t^2}{2})
\nonumber\\& \times \prod_{i=1}^{K}\Big[ 1 - \exp\Big(-\frac{(1+\sqrt{2(1+C)})t^2}{\beta_m{\overline {\gamma}_{{id}}}}\Big)\Big] 	
{\mathrm{d}}t\nonumber\\
&=\frac{\alpha_m}{2}\Big(1+\sum_{n=1}^{2^K-1}\frac{(-1)^{|\mathfrak{u}_n|}}{\sqrt{1+2\sum_{i \in \mathfrak{u}_n}\frac{1+\sqrt{2(1+C)}}{\beta_m\overline {\gamma}_{{id}}}}}\Big),
\end{align}
where the last expression follows from applying the binomial expansion and the fact $\int_0^{\infty}{e}^{-qx^2}dx=\frac{1}{2}\sqrt{\frac{\pi}{q}}$ {\cite[Eq. (3.318.3)]{table}. For a single-relay system, the average SER is
\begin{eqnarray}\label{SER_K1}
{\overline{P_s}}\hspace{-3mm} &=&\hspace{-3mm}\frac{\alpha_m}{2}\Big(1-\sqrt{\frac{\beta_m \overline{\gamma}_{{1d}}}{\beta_m \overline{\gamma}_{{1d}}+2+\sqrt{2}}}~\Big).
\end{eqnarray}
In the high SNR regime, i.e., $\bar{\gamma}_{id} \rightarrow\infty$, we have $1-\exp(-\frac{(1+\sqrt{2C})t^2}{\beta_m \overline{\gamma}_{{id}}})\approx \frac{(1+\sqrt{2C})t^2}{\beta_m \overline{\gamma}_{{id}}}$. Accordingly, the expression (\ref {SER00}) can be approximated as
\begin{eqnarray}\label{SER_fin}
{\overline{P_s^{\mathrm{DBCJ}}}}\hspace{-1mm} &\approx&\hspace{-1mm}\frac{\alpha_m\Big(1+\sqrt{2(1+C)}\Big)^K}{\sqrt{2\pi} \beta_m^K\prod_{i=1}^K {{\overline {\gamma}_{{id}}}}} \hspace{-1mm}\int_0^\infty\hspace{-2mm}{{t^{2k}} {e}^{-\frac{t^2}{2}} }{\mathrm{d}}t.
\end{eqnarray}
Using the facts that $\int_0^{\infty}x^{2k} {e}^{-p{x^2}}{\mathrm{d}}x=\frac{(2k-1)!!}{2(2p)^k}\sqrt{\frac{\pi}{p}}$ {\cite[Eq. (3.416.3)]{table} with $(2k-1)!!=\frac{(2k)!}{2^kk!}$, the high SNR approximation of the overall average SER can be obtained as
\begin{eqnarray}\label{SER_fin2}
{\overline{P_s^{\mathrm{DBCJ}}}}\hspace{-1mm} &\approx&\hspace{-1mm}\frac{\alpha_m (2K)!(1+\sqrt{2(1+C)})^K}{\beta_m^K 2^{K+1}K! \prod_{i=1}^K {{\overline {\gamma}_{{id}}}}}.
\end{eqnarray}
By inspecting (\ref {SER_fin2}), we observe that the DBCJ policy achieves a diversity order of $K$.
 
\section{Diversity-Multiplexing Tradeoff}
As is known, multi-antenna systems offer two different types of benefits in a fading channel: diversity order and multiplexing gain {\cite {Golds}}. The DMT describes this trade-off by presenting a trade-off between the error probability and the data rate of a system {\cite {Golds}}. Recently, the secrecy DMT of a
MIMO wiretap channel has been addressed in {\cite {rezki2}}, where a zero-forcing (ZF) transmit scheme is utilized.  In the following, we evaluate the DMT of the proposed JRP scheme from the perspectives of security and reliability.

\subsection{Secrecy Perspective}
Let $ d $ and $ r $ be the diversity order and the multiplexing gain of the presented system, defined as $d  \stackrel{\triangle}{=}  -\lim\limits_{\rho \to \infty}{\frac{\log_2 P_{\mathrm{out}}}{\log_2 \rho}}$ and $r \stackrel{\triangle}{=} \lim\limits_{\rho \to \infty}{\frac{{R_t}}{\log_2\rho}}$, respectively. Substituting (\ref {SOP_ap}) into the diversity order definition and using $R_t=r \log_2\rho$, the DMT can be obtained as
\begin{align}\label{rcl}
d&=\lim_{\rho \rightarrow \infty}{\hspace {-1mm}}
\frac{\sum_{i=1}^{K}\log_2{\overline{\gamma}_{{id}}}{\hspace {-.5mm}}{\hspace {-.5mm}}-{\hspace {-.51mm}}K\log_2{\hspace {-.51mm}}\Big(2^{2R_t}(1+\sqrt{2(1+C)}){\hspace {-.51mm}}-{\hspace {-.51mm}}1\Big){\hspace {-.51mm}} }{\log_2\rho}\nonumber\\
&=K(1-2r).
\end{align}
It is observed that when the multiplexing gain is not utilized ($ r \rightarrow 0 $), the diversity order equals the maximum value $d=K$, which is consistent with our previous results. On the other hand, a multiplexing gain of $ r = 0.5 $ is achieved as $d\rightarrow 0$. This is because, in the presented cooperative network, it takes two time slots to complete the transmission of one traffic flow, and hence, the maximum multiplexing gain of such network is just $0.5$. This can be dealt with utilizing a two-way network.\\

\subsection{Reliability Perspective}
In this case, $d  =  -\lim\limits_{\rho \to \infty}{\frac{\log_2 \overline{P_s}}{\log_2 \rho}}$. By substituting (\ref {SER_fin2}) into the diversity order definition and using $R_t=r \log_2\rho$, the DMT is given by
\begin{eqnarray}\label{dmt_r}
d=K(1-r).
\end{eqnarray}
One can conclude from (\ref {dmt_r}) that in the absence of a multiplexing gain ($ r \rightarrow 0 $), the diversity order is $K$. On the other hand, a multiplexing gain of $ r = 1 $ is achieved as $d\rightarrow 0$.

\section{Numerical Results and Discussion}
In this section, numerical examples are presented to verify the accuracy of the derived performance metrics (ESR, overall SOP and average SER) for the proposed JRP  transmission scheme. Numerical curves for the exact JRP scheme, marked with filled circles, are obtained using the optimum relay selection criterion in (\ref{selection_ex}) together with the OPA, which is numerically evaluated for a finite number
of source antennas using the bisection method. {We compare our proposed JRP scheme for both the NCE and CE cases with other well-known transmission schemes} as listed below:

 \begin{figure}[t]
 	\begin{center}
 		\includegraphics[width=3in]{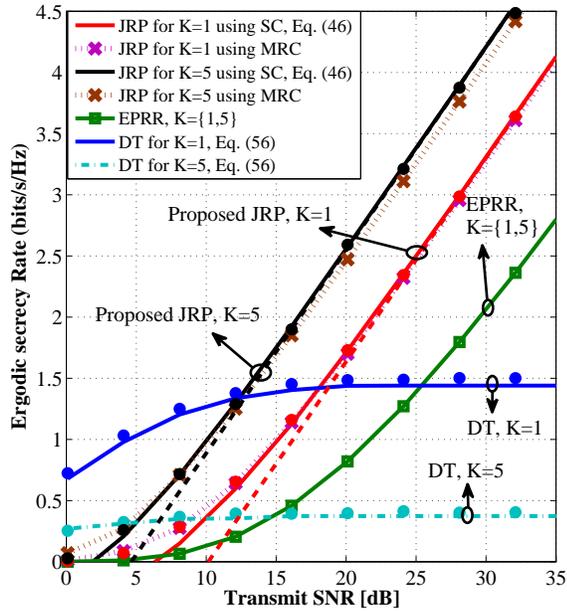} 
 		\caption{ESR versus the transmit SNR for the NCE case and for different transmission schemes. We consider $K=\{1, 5\}$ relays, the  multi-passive Eve $L=5$ scenario, and $N_\mathrm{s}=16$ source antennas. The asymptotic curves for the JRP scheme are shown with dashed lines using Eq. (\ref{rate-asymp-comp}), while the filled circles depict the Monte-Carlo simulations.}
 		\label{fig_ESR_SNR}\end{center}
 \end{figure}
 
 \begin{figure}[t]
 	\begin{center}
 		\includegraphics[width=3in]{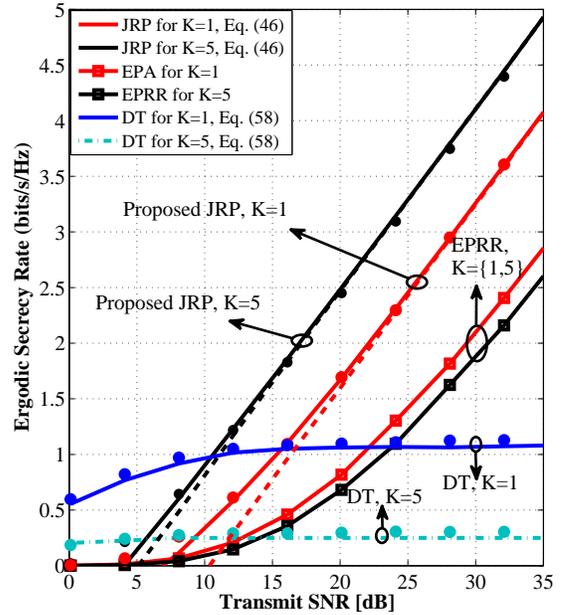} 
 		\caption{ESR versus the transmit SNR for the CE case and for different transmission schemes. We consider $K=\{1, 5\}$ relays, the  multi-passive Eve $L=5$ scenario, and $N_\mathrm{s}=16$ source antennas. The asymptotic curves for the JRP scheme are shown with dashed lines using Eq. (\ref{rate-asymp-comp}), while the filled circles depict the Monte-Carlo simulations.}
 		\label{fig_ESR_Ns_R5}\end{center}
 \end{figure}
 
\begin{enumerate}
	\item Equal power allocation (EPA) between source and destination ($\lambda^{NCE}=\lambda^{CE}=0.5$) with random relay selection, which is denoted by ``EPRR'',
	\item OPA between source and destination with random relay selection, which is denoted by ``OPRR'',
	\item EPA between source and destination with the optimum relay selection in (\ref{selection_ex}), which is denoted by ``EPRS'',
	\item The exact JRP scheme for the scenario of adopting the MRC technique at the malicious nodes for NCE case, and
	\item The DT policy, wherein both the untrusted relays and passive Eves are considered as pure Eves.
\end{enumerate}

To verify the accuracy of the derived LSMA-based expressions for the ESR of JRP, the SOP of JRP, the SER of JRP, the ESR of DT for NCE, the ESR of DT for CE, the SOP of DT for NCE, the SOP of DT for CE and the SER of JRP
 in (\ref{CapT_fin}), (\ref{SOP}), (\ref{SER00}), (\ref{dt_final}), (\ref{DT_ESR_CE}), (\ref{sop_dt}), (\ref{sop_dt_CE}) and (\ref{SER00}) respectively, we conduct Monte-Carlo simulations, where results are shown in Figs.
\ref{fig_ESR_SNR} - \ref{fig_ser}. Furthermore, we plot the derived asymptotic expressions for the ESR, SOP and SER of the proposed JRP scheme given by (\ref{rate-asymp-comp}), (\ref{SOP_ap}) and (\ref{SER_fin2}), respectively. As will be observed from Figs. \ref{fig_ESR_SNR} - \ref{fig_ser}, our LSAM-based closed-form expressions well matched with the simulation results, and the asymptotic curves well approximate the exact curves in
the high SNR regime. For simplicity and without loss of generality, we assume that the source, destination and relay(s) are located at the
positions (-1,0), (0,0), and (1,0) respectively, and the passive Eves are placed near the relays to overhear the maximum information. Unless otherwise stated, the values of network parameters are: number of antennas at the source $N_\mathrm{s}=16$, number of untrusted relays $K=\{1, 5\}$, number of passive Eves $L=5$, the target rate $R_t = 1$ $bits/s/Hz$ and the distance-dependent path loss factor $\alpha= 3$.

Figure \ref{fig_ESR_SNR} shows the achievable ESR versus the transmit SNR $\rho$ for NCE case. Our observations are summarized as follows:
\begin{enumerate}
	\item {The ESR performance of the proposed JRP scheme is a monotonically increasing function
		of the transmit SNR $\rho$ while the ESR  of the DT scheme converges to a constant value as presented in
		Section IV-B. The reason behind this behavior of the JRP scheme is that the injected jamming signal by the destination only degrades the received information signal at the malicious nodes and has no effect on the overall two-hop signal reception at the destination. As such, the secrecy performance of the DBCJ-based JRP scheme is superior to the performance of the DT policy in the high SNR regime, while
		the opposite behavior is observed at the low SNR regime. The reason is that by equipping the source with an LSMA and adopting an MRT beamformer the received SNR at the destination for DT becomes considerable, while the information leakage is negligible as computed in (\ref{g_DT}). We also mention that, as discussed in Section IV-A, the secrecy performance of the JRP is not satisfactory when the average transmit SNR of the second hop  is low. Therefore, the proposed JRP suffers from w secrecy performance loss in the low SNR regime.}
	
	\item The ESR performance of the JRP scheme increases as $K$ grows, which can be concluded  form (\ref{cap2}). This is because by increasing $K$, the probability of emerging a stronger channel between the relays and destination grows and accordingly, the secrecy rate increases.
	Contrary to the JRP scheme, the ESR of the DT policy decreases by increasing $K$. The reason is that according to (\ref{g_DT}), the received SNR at the destination is deterministic, while based on (\ref{leakage_NCE_DT}), the amount of information leakage increases as $K$ grows.
	\item  The secrecy rate advantage of JRP scheme compared with EPRR is obvious. For example, the SNR gap between the proposed JRP scheme  and EPRR is about 8 dB for $K=1$ case; this gap is about 13 dB for $K=5$ case to achieve the target transmission rate of 2 $bits/s/Hz$.
	
	\item The secrecy performance of the proposed JRP scheme with SC at the malicious nodes is close to the secrecy performance of the exact JRP with MRC technique.
\end{enumerate}
\begin{figure}[t]
	\begin{center}
		\includegraphics[width=3in]{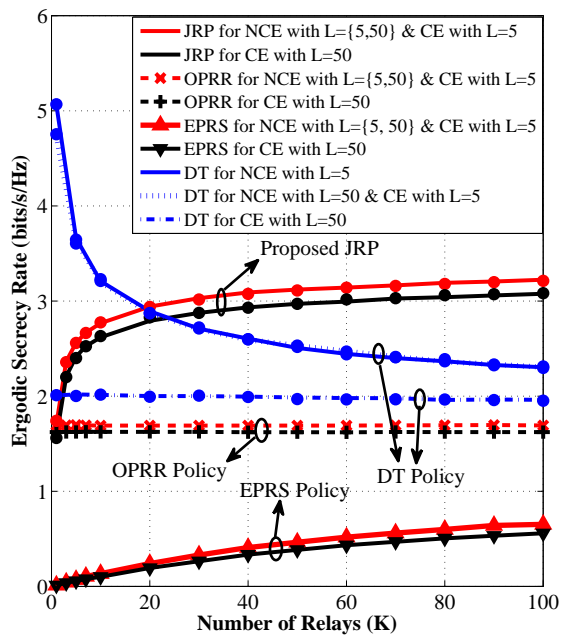} 
		\caption{ESR versus the number of untrusted relays for the transmit SNR of $\rho=20$ dB. We consider the multi-Eve scenario with $L=\{5, 50\}$, where the number of source antennas $N_\mathrm{s}=256$. The filled circles are obtained with the Monte-Carlo simulations.}
		\label{fig_ESR_relay}\end{center}
\end{figure}

In Fig. \ref{fig_ESR_Ns_R5}, we plot the ESR performance versus the transmit SNR $\rho$ for the CE case. As can be seen, the  proposed JRP scheme which is based on the near-optimal relay selection criterion in (\ref{selection})  is in perfect agreement with the exact numerical results across the entire SNR range of interest. As can be seen from Fig. \ref{fig_ESR_Ns_R5}, the ESR performance of the proposed JRP scheme increases by increasing $K$, while the opposite is observed for DT policy. Furthermore, we can conclude from Fig. \ref{fig_ESR_Ns_R5} that the ESR of the proposed scheme significantly outperforms the EPRR policy.

Figure \ref{fig_ESR_relay} examines the impact of the number of untrusted relays $K$ on the ESR performance of the proposed JRP scheme for the NCE and CE cases. We set the number of source antennas to $N_\mathrm{s}=256$ and the transmit SNR $\rho=20$ dB.  We also consider two cases of small $L=5$ and large $L=50$ number of passive Eves. The following observations can be made from Fig. \ref{fig_ESR_relay}:
 
\begin{enumerate}
\item As predicted by the analytical expressions and discussions in Section IV-B, the ESR performance of the proposed JRP scheme for NCE and CE cases is a monotonically increasing function of the number of untrusted relays $K$. This new result highlights that, unlike the results in {\cite{sun}}, {\cite{cioffi}}, the proposed LSMA-based scheme increases the  secrecy rate of the network by utilizing more untrusted relays.
\item The proposed JRP scheme for the NCE case with any number of Eves
offers a superior ESR performance relative to the CE case with large number of Eves $L=50$. This event can be justified simply based on the new results presented in Section III-B. The reason is that by equipping the source with an LSMA, the amount of information leaked to the malicious nodes in the first of transmission is negligible compared to that one in the second phase. Therefore, for CE with small number of Eves $L=5$, the ESR performance is the same as NCE case, while with large number of Eves, some information can be extracted by Eves, leading to ESR loss relative to NCE case. 
\begin{figure}[t]
	\begin{center}
		\includegraphics[width=3in]{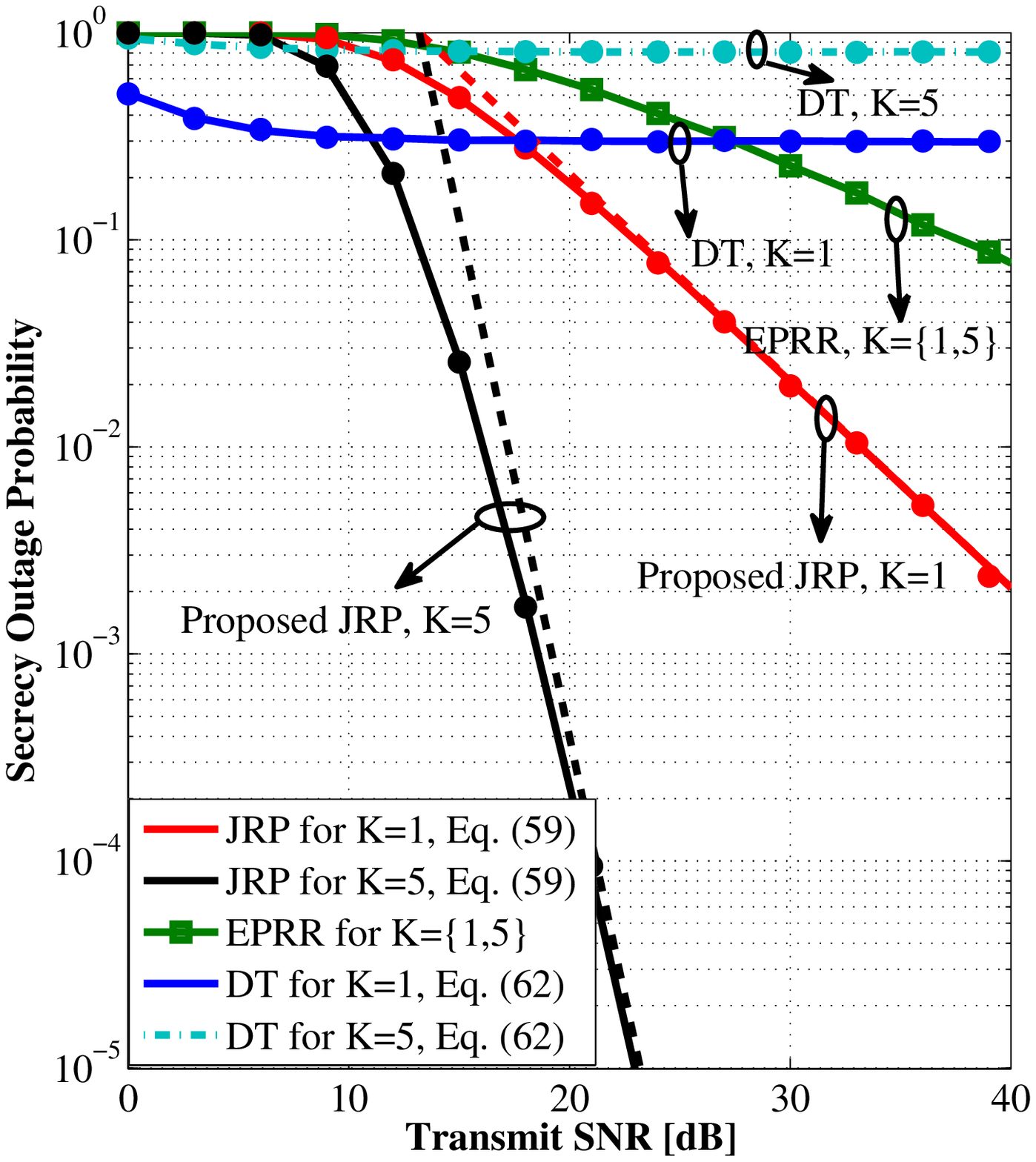} 
		\caption{Overall SOP versus the transmit SNR for the NCE case with $K=\{1, 5\}$ relays and $L=5$ passive Eves. The number of source antennas is set to $N_\mathrm{s}=16$ and the target rate to $R_t = 1$ $bits/s/Hz$. The asymptotic curves are shown with dashed lines using Eq. (\ref{SOP_ap}), while the filled circles depict the Monte-Carlo simulations.}
		\label{fig_sop}\end{center}
\end{figure}

\item When OPRR is adopted in the network, the ESR performance for both NCE and CE cases is a constant function of $K$. The reason is that 
 when one relay is randomly selected out of $K$ available relay nodes and OPA is applied, it is statically equivalent to the single-relay scenario with OPA. For large number of Eves $L=50$, due to the collaboration between Eves, the ESR of CE case is somewhat fewer than the NCE case.
\item The ESR performance of DT policy for NCE is a decreasing function of $K$. The reason is that by increasing $K$, the probability of emerging a stronger wiretap channel increases and therefore, the ESR decreases. However, for CE with large number of Eves, the ESR is a constant function with respect to $K$. The reason is that both the amount of information leakage and the received SNR at the destination are independent of the number of untrusted relays. We note that according to (\ref{leakage_NCE_DT}) and the law of large numbers, we have $\gamma_E^{CE}=\sum_{j=1}^L\gamma_j\approx L \rho \mu_{se}$ which is independent of $K$. 
\item The ESR performance of EPRS for both NCE and CE cases is an increasing function of $K$. This result highlights the effectiveness of relay selection in LSMA-based security networks even without applying OPA.
\end{enumerate}
\begin{figure}[t]
	\begin{center}
		\includegraphics[width=3in]{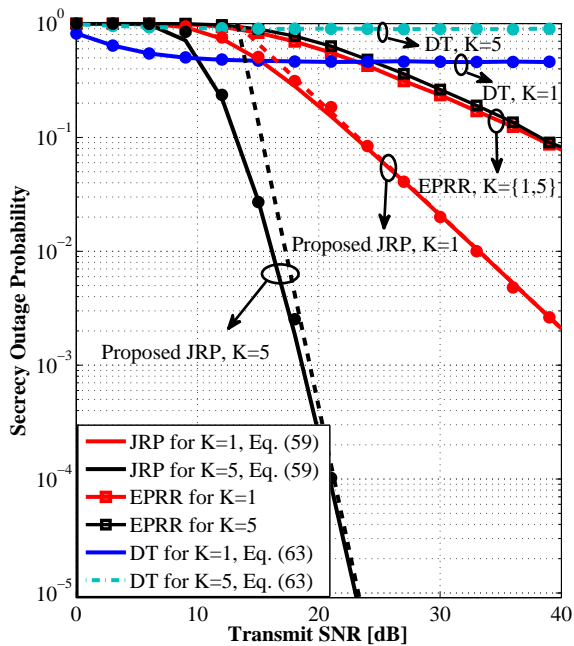} 
		\caption{Overall SOP versus the transmit SNR for the CE case with $K=\{1, 5\}$ relays and  $L=5$ passive Eves. The number of source antennas is set to $N_s=16$ and the target rate to $R_t = 1$ $bits/s/Hz$. The asymptotic curves are shown with dashed lines using Eq. (\ref{SOP_ap}), while the filled circles depict the Monte-Carlo simulations.}
		\label{fig_sop_R5}\end{center}
\end{figure}

\begin{figure}[t] 
	\begin{center}
		\includegraphics[width=3in]{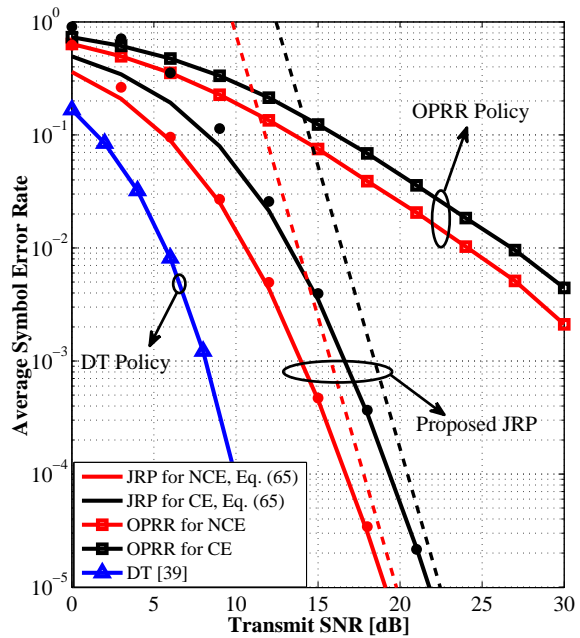} 
		\caption{Average SER versus the transmit SNR for the number of relays $K=5$ and multi-Eve $L=5$ scenario. The number of source antennas $N_\mathrm{s}=16$ and QPSK modulation is used. The asymptotic curves are shown with dashed lines using Eq. (\ref{SER_fin2}), while the filled circles depict the Monte-Carlo simulations.}
		\label{fig_ser}\end{center}
\end{figure}

In Figs. \ref{fig_sop} and  \ref{fig_sop_R5}, we compare the overall SOP versus the transmit SNR $\rho$ for NCE and CE cases, respectively. The results in Figs. \ref{fig_sop} and \ref{fig_sop_R5} indicate that the 
SOP performance advantage of our
proposed secure transmission scheme
compared to the EPRR policy. As can be readily observed from these figures, unlike DT policy that the SOP performance converges to a nonzero constant as $\rho \rightarrow \infty$, the SOP of the proposed JRP scheme for both NCE and CE cases approaches zero in the asymptotic SNR regime. Furthermore, as mentioned in Section IV-C, the proposed JRP scheme achieves the diversity order equal to the number of untrusted relays, which can be obtained simply from the asymptotic curves in Figs. \ref{fig_sop} and \ref{fig_sop_R5}. Finally, we observe from Fig. \ref{fig_sop_R5} that, while the DT policy fails to establish confidential communication for target secrecy rate of $R_t=1$ $bits/s/Hz$, the proposed JRP scheme enhances the PLS remarkably. This highlights the efficiency of the proposed JRP scheme.

To study the level of reliability of the proposed secure transmission scheme, we plot the average SER metric in Fig. \ref{fig_ser}. We consider the multi-relay scenario with $K=5$ and QPSK modulation. For this network topology, the DT policy offers a superior average SER relative to the proposed JRP scheme. The reason is that the DT policy provisions a better SNR at the destination compared with the proposed scheme. Furthermore, as observed, the average SER of NCE case is lower than the CE case. The reason is that according to the received SINR at the destination in (\ref{gamma_combine}), we have $\gamma_D^{CE}\leq \gamma_D^{NCE}$.
	Moreover, we can conclude from Fig. \ref{fig_ser} that the proposed JRP scheme is better compared to OPRR. For example, the SNR gap between the optimized network and OPRR
	is approximately 13 dB for NCE and 12 dB
	for CE, respectively, when  $\overline{P_s}=10^{-2}$. This  is because the proposed JRP  selects the relay with the largest second hop channel to signal transmission. Consequently, the achievable reliability of the proposed scheme is higher than the random relay selection of the OPRR policy. Evidently, unlike the OPRR, the proposed scheme attains diversity order $d=5$.

\section{Conclusion}\vspace{-1mm}
In this paper, we addressed secure transmission in a two-hop relaying network including a multiple antennas source, a single antenna destination, single antenna untrusted relays and single antenna passive Eves. We considered two practical scenarios of NCE and  CE. A novel JRP scheme has been proposed for security enhancement of the two NCE and CE networks.  For the proposed JRP scheme, a new closed-form expression was presented for the ESR  and the SOP as security metrics, and a new closed-form expression was derived for the average SER as a reliability measure over Rayleigh fading channels. We further evaluated the high signal-to-noise ratio slope and power offset of the ESR to reveal the impacts of system parameters on the achievable ESR. Our findings highlighted that the diversity order of the proposed JRP scheme is equal to the number of untrusted relays. Numerical results presented that the ESR of the proposed JRP scheme for NCE and CE cases increases as the number of untrustworthy relays grows. 

It would be interesting to extend the results in this paper to the case,
where trusted relays are also exist in the network. Thus, we
would like to consider this case as future work.

~~~~~~~~~~~~~~~~~~~~~{\bf Acknowledgments}\\
The authors deeply thank Prof. Dimitris Toumpakaris and Prof. Lajos Hanzo for their constructive comments to improve the paper. The authors also would like to thank the editor and the anonymous reviewers for the  constructive comments.

\end{document}